\documentclass[preprint]{aastex}
\usepackage{psfig}

\begin{document}

% GENERAL DEFINITIONS

\newcommand{\bq}{\begin{equation}}
\newcommand{\eq}{\end{equation}}
\newcommand{\kcorr}{$K$-correction}
\newcommand{\kcorrs}{$K$-corrections}
\newcommand{\dmm}{\mbox{$\Delta$m$_{15}(B)$}}

\bibliographystyle{apj}

\title{\kcorrs\ and Extinction Corrections for Type Ia
Supernovae}

\author{Peter Nugent, Alex Kim and Saul Perlmutter}  
 
\affil{Lawrence Berkeley National Laboratory, M.S. 50-F, 1
Cyclotron Road, Berkeley, CA 94720}

\email{e-mail: penugent@LBL.gov}

\begin{abstract}

The measurement of the cosmological parameters from Type Ia supernovae
hinges on our ability to compare nearby and distant supernovae
accurately. Here we present an advance on a method for performing
generalized \kcorrs\ for Type Ia supernovae which allows us to compare
these objects from the UV to near-IR over the redshift range
$0<z<2$. We discuss the errors currently associated with this method
and how future data can improve upon it significantly. We also examine
the effects of reddening on the \kcorrs\ and the light curves of Type
Ia supernovae. Finally, we provide a few examples of how these
techniques affect our current understanding of a sample of both nearby
and distant supernovae.

\end{abstract}

 
\section{Introduction}

In \citet{ham_kcor93} and Kim, Goobar and Perlmutter (1996)
\nocite{kim_kcorr96} (hereafter H93 and KGP96) the \kcorrs\ for samples
of Type~Ia supernovae (SNe~Ia) were presented at both low- and
high-redshift respectively. In H93 they dealt explicitly with the
standard single filter \kcorrs\ ($K_{BB}$ and $K_{VV}$).  KGP96
developed the ``cross-filter'' \kcorrs\ ($K_{BR}$ and $K_{VR}$)
necessary for the determination of the cosmological parameters from
high-redshift SNe~Ia. Cross-filter \kcorrs\ are necessary in the
high-redshift searches since it allows one to work with the rest-frame
optical light (where we have a good understanding of SNe~Ia) and
search photometrically where the observed light from a redshifted
SN~Ia is the brightest. Unlike the traditional single-filter \kcorr\
only a small amount of extrapolation over a given spectrum is
necessary. The only cost incurred by using cross-filter \kcorrs\ is
the small uncertainty in the zero points of the filter
system. Figure~\ref{kcorr_flt} shows both an ideal case (at $z=0.47$)
where the $B$- and $V$-band filters nicely match the de-redshifted
$R$- and $I$-band filters (and the cross-filter \kcorrs\ are
insensitive to the underlying spectrum) and a less desirable one (at
$z=0.65$) which involves some extrapolation or interpolation.

The dominant problem in applying \kcorrs\ on real data is that one
rarely has both photometric and spectroscopic observations of the same
supernova at the same phase of its evolution.  As a result, we are
obligated to resort to existing libraries of supernova spectra from
which we are to draw an appropriate \kcorr\ for the photometric
measurement in question.  The spectra and time-series evolution of
SNe~Ia are remarkably homogeneous \citep{1997ARA&A..35..309F} but
differences do exist \citep{nugseq95,bfn93} which are undoubtedly
linked to the variety of SN~Ia light-curve shapes and possible
foreground extinction.

The purpose of this paper is to further the work in KGP96 by
synthesizing our current understanding of the spectra and light curves
of SNe~Ia to provide the best possible method for determining their
\kcorrs\ along with their associated errors. In particular we
wish to focus on the effects that light-curve shape and reddening have
on the \kcorrs. One should note that the method presented here was
developed and used for the analysis of the 42 high-redshift supernovae
by the Supernova Cosmology Project (in \citet{42SNe_98}, hereafter
SCP99); it was also shared in advance of publication with the High-Z
Supernova Search Team (HZSST) for their analysis of 14 additional
high-redshift supernovae
\citep{riess_scoop98} and it was briefly discussed in $\S~5$ of
\citet{hdf_gnp99}. 

\section{The Problem}

The magnitude of a SN~Ia in filter $y$ can be expressed as the sum of
its absolute magnitude $M_x$, cross-filter \kcorr\ $K_{xy}$, distance
modulus $\mu$ and extinction due to dust in both the host galaxy,
$A_x$, and our galaxy, $A_y$.
\begin{eqnarray}
m_y(t(1+z)) = M_x(t,s) + K_{xy}(z,t,s,A_x,A_x)\nonumber \\
    + \mu(z,\Omega_M,\Omega_\Lambda,H_o) + A_x(t) + A_y(t).
\label{kcorr}
\end{eqnarray}
Here $t$ refers to the epoch when the SN~Ia is being observed, $s$ is
the stretch-factor [as described in \citet{7results97}, hereafter SCP97]
and $z$ is the redshift. The stretch-factor, $s$, simultaneously
parameterizes three aspects of the SN~Ia light curve. First, it is
responsible for changing the shape of the light curve. From an $s=1$
template, which is similar to the Leibundgut template for SNe~Ia
\citep{brunophd}, almost all known light curves for SNe~Ia can be reproduced
surprisingly well from the $U$-band through the $V$-band (over the
time range $-20 \lesssim t - t_0 \lesssim +40$) by stretching the time
axis of the light curve about maximum light by the factor $s$. This
fact is highlighted in Figure~\ref{phot_all} where we can see the
large reduction in photometric scatter after the application of the
stretch correction (see \citet{gold01} for a quantitative study of
this $B$-band stretch-factor fit). For the set of nearby supernovae in
Figure~\ref{phot_all} the peak-to-peak scatter for all photometric
bands about a $s=1$ template decreases by a factor of 3 and 2 for +15
and +40 days after maximum light respectively after the stretch
correction is applied to the data. Second, a relationship exists
between the peak magnitude in $B$ for a SN~Ia and $s$ that can be
expressed as follows (SCP99): \bq M_B(t=0,s) = M_B(t=0,s=1) -
\alpha*(s-1).
\label{alphaeq}
\eq 
This is very similar to the relationship found by \citet{phil93} where
a correlation between the peak brightness and the decline of the light
curve from maximum light to +15 days (\dmm) was presented. The broader
light curves (smaller \dmm) are intrinsically brighter, and the
narrower light curves (larger \dmm) are intrinsically fainter.
Finally, there is a relationship between the color of SNe~Ia and $s$
at peak $B$ brightness which can be expressed as follows:
\bq
B-V(t=0) = -\beta*(s-1) - \gamma.
\label{color_str}
\eq
This equation shows that the broader, brighter SNe are slightly bluer
as well. (Note that the best fit values for the constants in
Eqn's~\ref{alphaeq} \& ~\ref{color_str} and their uncertainties can be
found in SCP99. See also \citet{phillips99} for similar
relationships with respect to \dmm.)

The cross-filter \kcorrs\ presented in this paper were
calculated as they were in KGP96:
\begin{eqnarray}
  K_{xy}^{\rm counts} = -2.5 \log
    \left(
    \frac
       {\int \lambda{\cal Z}(\lambda)S_x(\lambda)d\lambda}
       {\int \lambda{\cal Z}(\lambda)S_y(\lambda)d\lambda}
    \right)
    +2.5 \log(1+z)\nonumber\\
   +2.5 \log
    \left( 
    \frac
        {\int \lambda F(\lambda)S_x(\lambda)d\lambda}
        {\int \lambda F(\lambda/(1+z))S_y(\lambda))d\lambda}
    \right).
\label{kceq}
\end{eqnarray}
Here $F(\lambda)$ is the spectral energy distribution (SED) at the
source (in this case a supernova), $S_x(\lambda)$ and $S_y(\lambda)$
are the effective transmission of the $x$ and $y$-bands and ${\cal
Z}(\lambda)$ is an idealized local stellar spectral energy
distribution for which $U=B=V=R=I=0$ in the photometric system used
here. $K_{xy}$ is defined as in Eq.~\ref{kcorr}. We have calculated
the \kcorrs\ as integrals of photon counts since here we have assumed
a count-based photometric system. For a full description of why a
count-based system is prefered see the appendix.

As can be seen in Eq.~\ref{kcorr} the \kcorrs\ are dependent on
anything which affects the observed spectrum. This not only includes
the obvious effects of redshift, $z$, and the spectral epoch, $t$, but
the effects of stretch, $s$, and extinction, $A_X$ and $A_Y$, as
well.\footnote{Note that a further complication, not considered in
this paper, would be for extinction along the line of sight that
occurred neither in our galaxy nor in the host galaxy of the
supernova.} As pointed out in \citet{nugseq95} individual spectral
features also vary slightly as a function of stretch in addition to
the overall change in the broad slope of the SED mentioned above. Thus
ideally what we need is a set of {\it template} SNe~Ia spectra that
cover a wide range in stretch, time and wavelength. Unfortunately at
this time such a set of spectra is not available. However, as we will
show, a very good approximation can be made that nicely accounts for
almost all the spectroscopic differences among SNe~Ia in order to
accurately reproduce their
\kcorrs.

\section{The Recipe}

We would like to develop a simple yet effective method for determining
\kcorrs\ that encompasses the effects of stretch and extinction,
overcomes gaps in existing spectroscopic data samples, and quantifies
the uncertainties. H93 noted the following characteristic of the
\kcorrs\ for SNe~Ia. ``One of the most interesting features of this
family of curves is their resemblance with the $B-V$ color curve of
Ne~Ia, implying that the $K$ term is basically driven by the color of
the SN.'' In Figure~\ref{clrcrv} we present the \kcorrs\ for a
typical, $s=1.0$, SN~Ia at a variety of redshifts along with its $B-V$
color curve. The similarity of these curves is striking and for very
good reason. It is the overall shape of the SED, i.e. the broad color
differences rather than the individual spectral line features, which
more than any other factor, determines the \kcorr. In this paper we
refer to these ``overall SED colors'' to describe the SED shape
averaged over $\sim$1000\AA\ bins.

Given this behavior we propose the following recipe for determining
the \kcorrs\ for a SN~Ia. 

\begin{enumerate}

\item Create a single template spectral grid of flux as a function of
wavelength and time for a $s \approx 1.0$ SN~Ia using all available
``Branch Normal'' SNe~Ia \citep{bfn93}.

\item Flux calibrate the spectra in the grid so that they yield as a
function of time the standard relative magnitudes (in $UBVRIJHK$) of a
$s=1.0$ SN~Ia.

\item Adjust the overall SED colors for each time step of the grid to
account for (a) extinction based on a given supernova's measured
$E_{B-V}$ and (b) the color differences as a function of the measured
stretch factor of a given supernova. Both of these color adjustments
are performed using a slope correction that follows the \citet{card89}
reddening law; as explained below, the stretch color differences have
a very similar wavelength dependence.

\item For a given SN~Ia measured properties ($s$, $z$, $E_{B-V}$)
calculate the \kcorr\ for each date, $t$,  using integrals over the
color-adjusted grid as in Eq.~\ref{kceq}.

\item Propagate uncertainties in $s$, $z$, $E_{B-V}$, $t$ and in the
constants of Eq~\ref{color_str} to estimate the \kcorr\ uncertainties.

\end{enumerate}

\section{Testing the Method}
\label{just}

Given the recipe outlined above we are now in a position to test the
validity of calculating \kcorrs\ from a single template of time
evolving spectra, where the SED of any given spectrum can be altered
to match the differences in stretch and/or extinction. Furthermore, we
need to test the effects of interpolation in time between samples of
observed spectra. While the database of SNe~Ia has grown considerably
in the past few years, there are very few spectra at a given epoch for
a given supernova that go from the UV through IR. Additionally, for a
given supernova, there are often large gaps in the spectral sequence,
brought about by the supernova going behind the sun, a full moon or
lack of continuing access to telescope time.

We have made a couple of hypotheses here that we must test.  Do
different SNe~Ia ($s$, $E(B-V)$) at the same epoch have the same
\kcorr\ after a color-adjustment on the overall SED color is applied
(step 3 above)?  We test this by examining the \kcorrs\ for two very
different SNe~Ia 1992A and 1991T. The stretch values for these
supernovae are nearly at opposite ends of the observed range, 0.80 and
1.09, and their reddening corrected $B-V$ colors at maximum light were
-0.01 and -0.07 respectively (SN~1991T suffered from an extinction of
$E(B-V) \approx 0.2$). In Figure~\ref{kc_str}a we see $K_{BR}$ at $z =
0.35$ for both these supernovae and we note differences of at least
0.1 to 0.2 mag. over this time range. We chose this redshift, which
involves a fair amount of extrapolation between the rest-frame
$B$-band and de-redshifted $R$-band filters, to highlight the
differences between the two SNe. In Figure~\ref{kc_str}b we have
applied a slope correction to each of the supernova's spectra so that
they have the same $B-V$ colors as an $s = 1.0$ SN~Ia at every
epoch. The slope correction was performed by altering the flux using
the reddening law of \citet{card89} (either making them bluer or
redder accordingly). This is a reasonable approach since, while not
exact, \citet{vdb95}, \citet{rpk96dust} and \citet{trpnbrn99} have
noticed that the intrinsic relationship between $M_V$ and $B-V$ for
SNe~Ia of varying stretch is quite similar to that of a typical dust
reddening law. The result of these ``color corrections'' to the
spectra is that the
\kcorrs\ now lie on top of each other with differences between the
curves of at most 0.02 mag. It is well known that these particular
supernovae show several differences spectroscopically (e.g. see
\citet{nugseq95} and \citet{bfn93}), however, for the \kcorrs\ these
differences are quite small compared to the effect of the
overall SED colors.

We do note the following exception to this method. SN~1991bg, a
peculiar, sub-luminous SN~Ia (with $s<0.7$), has a large trough in its
spectra due to several strong \ion{Ti}{2} spectral lines
\citep{fil91bg}. This feature is over 400\AA\ wide and persists quite
strongly for several months. The color difference between a $s=1.0$
SN~Ia and a 91bg-like supernovae at maximum light is $> 0.8$ in $B-V$,
and is due almost completely to the presence of these \ion{Ti}{2}
lines. The ``color correction'' method mentioned above will not work
with these types of SNe~Ia. Instead, one would have to create a set of
template spectra based solely on 91bg-like supernovae. 

How well do the temporal interpolations work?  Since we see that it is
the overall SED colors which are the largest factors in the \kcorr,
and not individual spectral features, we have some confidence in
interpolating between different supernovae across a wide range of
epochs.  A test of this method is seen in Figure~\ref{kc_time}. Here
we once again display $K_{BR}$ for a $z=0.35$ SN~Ia with $s=1.0$ along
with three other sets of \kcorrs, chosen to mimic the effect of
mis-identifying the date of a spectrum or poorly interpolating between
dates. These \kcorrs\ were calculated by forcing the spectra within 10
days of a given point on the light curve to have the same colors as
the SN~Ia has at that central point. This was done for maximum light,
$+10$ days and $+20$ days. The difference in the \kcorrs\ is, at most,
0.05 mag over the entire 20 day window and $\lesssim$ 0.02 mag. for a
10 day window. Thus the spectral feature evolution of a SN~Ia has a
negligible contribution to the change in the \kcorr\ over periods of
roughly one week compared to the changes in the overall SED
colors. This also gives us confidence in interpolating between spectra
of slightly separated epochs, so long as the overall SED colors are
correct.

\section{SN~Ia Template}
\label{template}

Table~\ref{templ_tab} lists the SNe~Ia used to construct the
template. In Table~\ref{stretch_col} we list the $UBVRIJHK$ magnitudes
used to determine the overall SED colors for this $s=1.0$ SN~Ia
template. Note that this is a preliminary template and will certainly
change in the future as more and better observations of SNe~Ia are
obtained. {\it The preliminary nature of this template has no effect
on the results presented in this paper, since our goal here is to
stress the method rather than the actual values for the \kcorrs}. The
color template was created from four primary sources. The $B$- and
$V$-band light curves come from the data of \citet{riess_rise99} and
the $R$, $I$ light curves from \citet{rknop_baas02} (see also
\citet{gold01} for comparable $B$-band templates). . The $JHK$
photometry comes from the work of \citet{elias_jhk81} and the $U$-band
data from the work in \citet{branetal96}. The early, rising, portion
of the light curve, which is well mapped out in $B$ and $V$ prior to
-10 days, was extrapolated in all other filters in a smooth way such
that the explosion date for this SN~Ia occurs 20 days before maximum
light in $B$.

The set of observed spectra was assembled in a two dimensional grid of
flux as a function of time (1 day bins with $-19 < t < 70$) and
wavelength (10 \AA\ bins with 2500~\AA\ $ < \lambda < $ 25000~\AA). If
two or more spectra on a given day overlapped we used their
noise-weighted average. All the spectra were then ``color corrected''
by the method outlined above to have the colors listed in
Table~\ref{stretch_col}.  To fill in the temporal gaps, a simple
linear interpolation was applied throughout the set. Extrapolation was
only used prior to day $-14$ and just for the UV after day $+45$. The
entire set was then smoothed in a given wavelength bin, using a
gaussian-weighted boxcar approach, over time (a $\sigma$ = 2 days was
used). This removed any ``glitches'' (such as poor cosmic-ray
subtractions, etc.) from the overall template. Finally a last ``color
correction'' was applied to this grid of spectra.\footnote{Note that
since there is so little spectroscopy available for $ \lambda >$
10000~\AA\ for SNe~Ia we just assumed monotonically decreasing fluxes
as a function of wavelength in these regions which reproduced the
photometry. We included this region of the spectrum purely for
convenience in preparation of future work on bolometric light curves.}
In Figure~\ref{spectra} we present a sampling of the template from two
weeks before to four weeks after maximum light.

\section{Error Budget}

We now turn our attention to the error associated with making a given
\kcorr. We have already considered the first component of this error
in \S~\ref{just}. This is the error incurred when using the template
method and is a result of the effectiveness of the ``color
correction'' across the SN~Ia sub-type and the interpolation in time of
the spectra. In the worst-case-scenario for rest-frame optical
cross-filter
\kcorrs\ ($z \approx 0.35$) there is $<$ 1\% {\it rms}
dispersion over the entire light curve due to differences in supernova
sub-type, and $<$ 1\% dispersion due to interpolations in time of the
spectra, when constrained to $\pm$~4 days.

The second component of the error is purely observational and depends
on two factors: (1) the alignment of the de-redshifted filter (on
which the \kcorr\ will be performed) with respect to the rest-frame
filter and (2) the photometric error on the observed color of the
supernova. This overall observational error vanishes for either
perfectly matched filters or zero error in the color measurement.

To illustrate this error we present in Figure~\ref{kcorr_err} $K_{BR}$
as a function of $z$. Here we have calculated this \kcorr\ for three
different cases in which we assign an uncertainty in $R-I$ of 0.02,
0.05 and 0.10. The most striking fact about this graph is how the
uncertainty in $K_{BR}$ vanishes at $z = 0.49$, where the rest-frame
$B$ and de-redshifted $R$-band filters nicely overlap, regardless of
the uncertainty in the color measurement. One also notices how the
uncertainty in the \kcorr\ flares out at the upper and lower ranges in
redshift. In practice, when trying to obtain rest-frame corrected
$B$-band photometry, $K_{BR}$ is not used for SNe~Ia with $z
\lesssim 0.3$ or $z \gtrsim 0.7$. A switch is made to $K_{BV}$ or
$K_{BI}$, respectively, to perform the \kcorr. At these redshifts,
where the rest-frame $B$ and de-redshifted $R$-band filters barely
overlap, the uncertainty in the observed color of the supernova
translates directly into a larger uncertainty in the \kcorr. For
example, at $z=0.3$ we note that $\sigma_{K_{BR}} \approx
\sigma_{B-V}$. While this may seem like it would impart a large
uncertainty in the corrected $B$-band magnitude it is in fact dwarfed
by the error in the extinction correction. Given that $A_B = 4.1
E(B-V)$ we see the fact that $\sigma_{A_B}$ is a factor of four larger
than the $\sigma_{K_{BR}}$. {\it Thus the major problem facing
individual high-redshift SN~Ia photometry will always lie in the
accuracy of multi-color photometry, not in performing \kcorrs.}

\subsection{Problems with Restframe $U$-band Photometry}
As automated searches for supernovae continue to progress, one of the
greatest problems faced in the reduction of this data concerns our
knowledge of the rest-frame UV light. Many problems exist with current
$U$-band photometry from an observational standpoint alone
\citep{richsn94d}. When this is coupled with the large intrinsic
variations among similar SNe~Ia in the UV \citep{branetal96} and the
larger uncertainties in absolute UV photometry due to extinction, one
can easily see that that working in this area of the spectrum is
problematic at best. Here we provide two examples of how the above
stated problems affect our current understanding of SNe~Ia.

The highest redshift SNe~Ia that have been confirmed spectroscopically
are SN~1998eq \citep{albinoni_iau} at a $z=1.20$ and SN~1999fv
\citep{tonry_spe} at a $z=1.19 \pm 0.02$. Both supernovae were discovered in
the $I$-band. The SCP was able to follow SN~1998eq in the $J$-band
with NICMOS on board the {\it HST} while the HZSST obtained a deep
image of SN~1999fv in the $J$-band from the ground
\citep{tonry00}. Both teams will rely on the $I$-band photometry of
the supernovae to determine the shape of its light curve (the $I$-band
maps almost directly with the rest-frame $U$-band at these redshifts)
and use the $J$-band photometry to infer the peak rest-frame $B$-band
magnitude. However, current uncertainties in the intrinsic $U-B$
colors of SNe~Ia are quite large. A $s=1.0$ SNe~Ia has $\sigma_{U-B}
\ge 0.1$ which implies an uncertainty in $A_B \ge 0.5$ ($A_B = 4.8
E_{U-B}$)
\citep{branetal96,nature98}. Without any other constraints on the
total reddening these supernovae become quite ineffective in
discriminating between various cosmological models.

Not only does our current understanding of the UV affect our
interpretation of the very high-redshift SNe~Ia, but it affects our
knowledge of nearby supernova and therefore our current understanding
of the relationship between light-curve shape and peak brightness as
well. In Table~\ref{ulow} we present the differences between SN~1992A
and SN~1994D in their rest-frame $U-B$ and $B-V$ colors and in $K_{BB}$
if these SNe~Ia were found at a redshift of $z=0.1$. These two SNe~Ia
have very similar stretches ($s \approx 0.81$) but very different
colors in the blue as is seen quite clearly in Figure~4 of
\citet{patat94d}. Table~\ref{ulow} highlights how these differences
lead to very different \kcorrs\ as a function of time. Depending on
the sampling, a SN~Ia observed at this redshift whose \kcorr\ did not
account for color differences in the UV could yield reduced $B$-band
light-curve shapes which are incorrect by $>$10\%. This difference
would create a 0.17 magnitude offset in the distance modulus (see
Eq.~\ref{alphaeq}). SNe~Ia at this redshift are particularly important
for the Hubble diagram since they are well out in the Hubble flow and
thus unaffected by host galaxy peculiar velocities. One possible
explanation of the difference in the UV color can be found in the work
of \citet{hof_q098} where the metallicity of the progenitor can have a
strong influence on the UV spectrum. \citet{lentz99} has quantified
these effects by varying the metallicity in the unburned layers and
computing their resultant spectra at maximum light with the spectrum
synthesis code PHOENIX \citep{nuge_hyd97,nugphd}. The differences seen
in the theoretical calculations easily span the range seen between
SNe~1994D and 1992A and are quite plausible as an explanation for the
large intrinsic scatter in the UV for SNe~Ia given the range in
progenitor environments for these objects \citep{hamuy_env00}.

We conclude this section with a final note on how the intrinsic
differences in the $U$-band photometry can affect a bias on the
high-redshift search for SNe~Ia. Given the differences seen in the
$U-B$ colors for the two SNe~Ia above, we can perform a simple
Monte-Carlo search for supernovae along the lines of
\citet{hdf_gnp99}, to estimate the likelihood of finding the bluer
SN~1994D's over the redder SN 1992A's. In
Figures~\ref{search_r},~\ref{search_i}~and~\ref{search_ratio} we have
plotted the discovery rate for a search with a 50-50 split between
supernovae like 1994D and those like 1992A. In the $R$-band search we
have modeled the Monte Carlo after a typical 4-m telescope run like
those carried out by the SCP (SCP99) and HZSST
\citep{riess_scoop98} between 1995 and 1998. The search covers 4
square degrees with a 28 day separation between the reference and
discovery images and has limiting magnitudes, typical for those
searches, of 23.0, 23.5 and 24.0. What is clearly seen from the
figures is the strong bias above a redshift of 0.5 (where the
restframe $U$-band light slips into the $R$-band filter) for
discovering blue SNe~Ia. Of the SNe Ia presented in the aforementioned
papers over 33\% are at $z > 0.5 $ and thus susceptible to this bias.
More recently the searches conducted by both teams have been carried
out in the $I$-band with the goal of discovering even higher redshift
supernovae. For this Monte Carlo we assumed a search gap of 28 days
and 1 square degree to limiting magnitudes of 24.0, 24.5 and
25.0. Going to the redder filter we see the bias affecting the search
for $z > 0.9$.

How would such an effect alter the measurement of the cosmological
parameters? First one would have to know the relative rates of
94D-like to 92A-like supernovae at any given redshift which is
completely uncertain at the moment. However, even assuming an equal
distribution of the rate of both populations, on face value there
would be little difference. Both SN 1992A and SN 1994D appear to have
similar absolute peak brightnesses in the restframe $B$-band given the
intrinsic uncertainties in the SN~Ia calibration and the distances to
their host galaxies \citep{ajhar}. The major difference would lie in
color corrections based on the uncertainties in the \kcorrs\ for those
supernovae at redshifts where the restframe $U$-band light
contaminates the correction to the $B$-band measurement. Given a split
distribution and the Monte Carlo simulations above we would expect to
find, on average, bluer supernovae at higher redshift and potentially
underestimate the correction due to host galaxy extinction. This would
lead one to overestimate distances to these objects.

However, in general, this effect is quite limited. This is due to the
fact that the total amount of reddening a supernovae can have which is
discovered near the edge of detection is quite small. This is further
amplified by the fact that since one is looking at the rest-frame UV
light, which suffers a higher relative extinction, even less reddening
is allowed before the supernova becomes too extinguished to
discover. Even for those 94D-like supernovae with a small amount of
extinction that still get discovered, the effect of under-correcting
for extinction, and thus overestimating the distance to those
supernovae, is compensated by the fact that these supernovae are
brighter in the UV and the flux in the observed filter is higher than
average.

In practice, as was reported in \citet{rknop_baas02}, one can use both
the 94D-like templates and the 92A-like templates to compute the
\kcorrs\ for the high redshift supernovae. The differences in the
overall fits to the cosmological parameters are quite small, though
one definitely produces systematic differences in the color
distributions of the high-redshift supernovae based on the choice of
template. The simple solution to this problem is to observe the
high-redshift SNe~Ia farther into the red where the restframe $U$-band
light does not play a significant role in determining the corrected
$B$-band magnitude.

\subsection{Problems with Restframe $I$-band Photometry}

One of the interesting characteristics of SNe~Ia is the shape of the
light curves in the near IR. In the $I$-band there exists a secondary
peak which occurs more strongly, and at a slightly later epoch, for
the more luminous supernovae while less luminous SNe~Ia like SN~1991bg
hardly show this secondary peak at all (e.g. see
\citet{hametal96,riess_data99,kevinir_01}). This feature in the
$I$-band light curves is in part due to the presence of the strong
emission from the Ca~II IR triplet. The emission is delayed in the
brighter supernovae and is much stronger when it peaks compared to the
fainter supernovae.

In the restframe, the $I$-band straddles this P-Cygni profile which is
almost 1000~\AA\ wide as seen in Figure~\ref{iband}. Even for low-$z$
supernovae this feature poses a problem. At a redshift of just a few
percent ($z \approx 0.05$) the $I$-band filter moves completely off
the emission feature. Thus \kcorrs\ which are based on the template
method outlined above which can not compensate for the change in this
feature's strength due to epoch and luminosity-class can be in
significant error. In \citet{lou99ac_02} differences approaching 0.3
magnitudes were seen using various template methods compared to using
the actual supernovae spectra to perform the \kcorr. Clearly the
solution to this problem is to derive separate templates based on
luminosity class for performing the \kcorrs. This is exactly the same
type of problem mentioned before in the $B$-band for the
low-luminosity SNe~Ia due to the presence of the Ti~II trough.

\section{Reddening}

We now turn our attention to the effects of reddening on the light
curves of SNe~Ia. There are two factors we must consider here. First,
as pointed out by KGP96, the observed color excess of a SN~Ia will
vary slightly with time because the rapidly varying SED shifts the
effective wavelength of each filter. Second, \citet{phillips99}
(hereafter P99) have noted that the observed decline rate of a SN~Ia
is a weak function of the dust extinction which affects the light
curves. Both of these effects directly impact one's ability to
accurately measure the light-curve shape and peak corrected-magnitude
of a SN~Ia.
 
We start with a few definitions. $E(B-V)_{true}$ is a measure of the
absorbing gas and dust along the line of sight to the supernova, not
the observed color excess, $E(B-V)_{obs}$, which will vary with
supernova epoch and total extinction. In Figure~\ref{red_time} we have
plotted the relative variations of
\begin{eqnarray}
R_{X} = A_{X}/E(B-V)
\end{eqnarray}
as a function of time. The calculations were performed for $B$, $R$
and $I$ ($V = R_{B} - 1.0$ by definition and thus is not shown) using
the spectral template of \S~\ref{template}. We display both
$R_{B_{obs}}$ and $R_{B_{true}}$. Each serves a purpose for performing
color corrections depending on whether one wants to correct photometry
based on their observed colors, or if one {\it a priori} knows
$E(B-V)_{true}$ along the line of sight to the SN~Ia in question. Over
the first 40 days of the light curve $R_{B}$ is found to vary
considerably, $-0.1 \lesssim R_{B} \lesssim 0.4$. The variation is
caused by the rapid evolution in the spectra of SNe~Ia around maximum
light noted by the change in $\lambda_{eff}$ as mentioned above.

P99 found the following relationship between the {\it true} decline
rate and the {\it observed} decline rate as a function of
$E(B-V)_{true}$.
\begin{eqnarray}
\dmm_{true} \simeq \dmm_{obs}+0.1E(B-V)_{true}
\label{redeq}
\end{eqnarray}
This shows that from peak to +15 days the light curves of SNe~Ia get
{\it broader} as they become more extinguished by dust. In
Figure~\ref{red_lcs} we plot the relative difference in $B$-band
magnitude (normalized at maximum light) between an unextinguished
light curve and several increasingly extinguished light curves. These
curves were calculated using the spectral template from
\S~\ref{template} and the same reddening law employed by P99
\citep{card89}. Here one can clearly see the result found by P99,
however, the pre-maximum portion of the light curve does not behave in
a similar way. Our results show that extinguished light curves are
narrower prior to maximum light and broader past maximum light. The
explanation for this result is quite simple. For a given epoch, a
reddened spectrum produces a shift in the effective wavelength,
$\lambda_{eff}$, of the spectrum, $F(\lambda)$, integrated through the
filter, $S(\lambda)$, toward the red. $\lambda_{eff}$ is given by the
following equation:
\begin{eqnarray}
\lambda_{eff} =  \frac{\int \lambda F(\lambda)S(\lambda)d\lambda}
        {\int F(\lambda)S(\lambda)d\lambda}.
\end{eqnarray}
In Figure~\ref{lameff} we have plotted $\lambda_{eff}$ as a function
of time for various levels of extinction in the $B$-band. We note that
a supernova naturally becomes cooler and redder as a function of time
after maximum light and for a given amount of extinction there is also
the expected shift in the effective wavelength. Thus a simple minded
explanation for the change in a heavily reddened SN~Ia's $B$-band
light curve is that the shift in effective wavelength makes the light
curve act more like a $V$-band light curve, which is broader after
maximum light but narrower prior to it (see
Table~\ref{stretch_col}). Therefore, it would be incorrect to
interpret the equation in P99 as something which is applied to a given
\dmm light-curve class. Rather it should only be used in the strict
interpretation of \dmm, the change in magnitude from peak to +15 days.

\section{Examples of Extinction and \kcorrs on Observed Data}

In this section we look at the application of the template method with
regards to both reddening corrections on three nearby SNe~Ia and \kcorrs\
on one of the SCP's first set of seven high-redshift supernovae. 

\subsection{SNe 1998bu, 1986G and 1996ai}

SNe 1998bu \citep{sunetal99,jha99_98bu}, 1986G \citep{phil86g} and
1996ai \citep{riess_data99} were three well observed SNe~Ia caught
prior to maximum light and followed regularly for over two months in
multiple filters. All three suffered from a large amount of
extinction: SN 1998bu had $E(B-V) \approx 0.37$, SN 1986G had $E(B-V)
\approx 0.6$ and SN 1996ai had $E(B-V) \approx 1.68$.\footnote{Here the
values for the extinction were determined via a fitting method which
incorporates the spectral template mentioned in \S~\ref{template}.
Within their corresponding uncertainties, they agree with the values
published in the photometry papers mentioned above.}

As outlined in the previous section, the biggest effect that the
reddening has on a SN~Ia (outside of the obvious color change) is its
ability to alter the observed stretch of the light curve. For these
three SNe the $B$-band stretch should be {\it decreased} from the
observed values by the following amounts: SN~1998bu -- 3\%, SN~1986G
-- 5\% and SN~1996ai -- 11\%.  Eq.~\ref{alphaeq} implies that without
performing the aforementioned correction to the stretch due to the
extinction along the line of sight to the supernovae their distance
moduli would be overestimated by 0.05, 0.09 and 0.19 mag respectively.

A secondary effect one needs to consider is that a given stretch
implies a particular $B-V$ color evolution for the supernova. If one
used the peak observed $B-V$ color of these supernovae coupled with
their observed $B$-band stretch they would falsely arrive at a higher
value of $E_{B-V}$ for each. This is simply due to the fact that the
observed $B$-band stretch of an extinguished SN~Ia is broader than its
true, intrinsic stretch and a narrower stretch SN~Ia is redder than a
broader one. Therefore using the stretch corrections to the
extinguished SNe above coupled with Eq.~\ref{color_str} we see that
the true value of $A_B$ would be altered by the following amounts:
SN~1998bu -- 0.02, SN~1986G -- 0.04 and SN~1996ai -- 0.09 mag. This
goes in the opposite direction of the corrections outlined above and
thus slightly mitigates the overall effect that reddening has on the
distance determinations to SNe~Ia.  However ignoring this effect
increases the overall dispersion in the stretch luminosity
relationship.

\subsection{SN 1994H}

As seen in SCP99, SN~1994H (one of the first seven supernovae
published in SCP97) was one of two large outliers in the Hubble
diagram of the SCP's high-redshift SNe~Ia. It was also one of the few
supernovae without a spectrum to confirm its classification and it was
excluded from all of the primary fits of SCP99. Using a brute-force
multi-dimensional $\chi^2$ fitting program which incorporates the
template in \S~\ref{template}, we have calculated the best fit to the
light curve of this supernova (a modified version of this method was
used in SCP99. From this we find the best fit stretch to be
$s=0.90\pm0.06$. The supernova had an observed color which was bluer
than $B-R=0.7$ at 0.0$\pm$0.6 days at the 95\% confidence limit. A
SN~Ia at this redshift ($z=0.374$) and stretch will have an
unextinguished $B-R$ color of 1.63$\pm$0.05 (the error bars are the
result of the uncertainties in the time of maximum light and
stretch). The difference between the observed color and the one
calculated for a SN~Ia of this stretch are many standard deviations
apart. Reddening, of course, only makes this supernova more
discrepant.

This result, seen in SCP99, differs from the result first
presented in SCP97 due to the fact that the fitting procedure used at
that time did not vary the \kcorrs\ as a function of stretch. At this
redshift, where the extrapolation is large, these differences are
considerable. For SCP97 the stretch was determined to be 1.09$\pm$0.05,
which would place this supernova in the very broad--bright--blue
category of SN~Ia along the lines of SN~1991T\citep{phil91t}. The
$B-R$ color of a SN~1991T-like supernova at this redshift is $B-R
\approx 1.0\pm 0.15$, which, as pointed out in SCP97, is only in mild
disagreement with the observed color.\footnote{In addition, one should
note that improved final-reference photometry taken of SN~1994H after
SCP97 was published produced small changes in the overall light-curve
shape.}.

The implication of this result is that SN~1994H was most likely a Type
IIL or II-n supernova caught at an early phase. Given that most SNe~II
look like blackbodies early in their light curves, the $B-R$ color of
0.7 mag is not too surprising. In fact, the color at this redshift
implies a blackbody temperature of 9750 $^{\circ}$K, which is quite
consistent with a SN~II at this phase. Given a cosmology of $h=0.635$,
$\Omega_M=0.3$ and $\Omega_{\Lambda}=0.7$ with a $K_{BR}=-0.68$ for a
9750 $^{\circ}$K blackbody, we find that SN~1994H would have an $M_B
\gtrapprox -19.60$. This would make it a rather luminous SN~II along
the lines of SN~1979C \citep{branepm,sn79cdev}. This result
strengthens the case for the primary cosmology fit of SCP99 which
excluded SN 1994H.

\section{Conclusions}

We have presented a method for performing generalized \kcorrs\ for
SNe~Ia which allows us to compare these objects from the UV to near-IR
over the redshift range $0.0 < z < 2.0$ along with the corresponding
uncertainties currently associated with this method. These
uncertainties are in general much smaller when compared to the
uncertainties in the extinction corrections. Future data will
certainly improve upon this method considerably, especially if
properly calibrated spectral templates can be acquired for each
sub-class of SNe~Ia. We have also examined the effects of reddening on
the light curves of nearby SNe~Ia and how our current, poor
understanding of the behavior of SNe~Ia in the UV effect the use of
SNe~Ia at both high- and low-redshift.

\acknowledgments 
We would like to thank Mark Phillips for many helpful discussions,
particularly those concerning the $U$- and $I$-band photometric
measurements of nearby SNe~Ia. We would also like to thank Adam Riess
and Weidong Li for a discussion that led to a great simplification of
the section on extinction corrections and Robert Knop on supplying
$R$- and $I$-band templates of nearby SNe~Ia prior to
publication. This work was supported by a NASA LTSA grant to PEN and
by the Director, Office of Science under U.S. Department of Energy
Contract No. DE-AC03-76SF00098. AGK acknowledges support from NSF
grant 21434-13066.

\appendix 
\label{app1}

\centerline{APPENDIX}

\section{Photon versus Energy K-corrections}

Measurements of the cosmological parameters using distance indicators
rely on the redshift-dependent evolution of the distance modulus
$\mu$.  The distance modulus is measured as the difference between
observed and absolute magnitudes of a ``standard candle'' after
\kcorr\ \citep{ok:kcorr} for the redshifting of its spectrum.
The theoretical value for $\mu$ is related to the luminosity distance
$d_L(z)$ defined such that a source with luminosity $L$ at redshift
$z$ has observed energy flux $f$ as if the energy has been diluted to
the surface of a sphere with radius $d_L$, i.e.  $L=4 \pi d_L^2 f$
(e.g. \citet{carrollpressturner}).  Cosmological parameters can then
be measured from their functional dependence on $d_L$.

Observations are in fact made with photon counters (CCD's,
photo-multipliers) and the luminosity distance is not the same as the
``photon luminosity distance'' $d_\gamma$; if N is the photon
luminosity and n is the observed photon flux, then $N = 4 \pi
d_\gamma^2 n$ where $d_L=d_\gamma (1+z)^{1/2}$.  This has lead to some
confusion as to whether a ``photon'' distance modulus should be used
to measure cosmological parameters, whether the magnitude system is
photon-based or energy-based, and which \kcorrs\ should be
applied.  Such distinctions which previously have been unimportant are
significant as we move into an era of precision cosmology.  In this
appendix we rederive and expand upon the \kcorr\ results of
\citet{1983ApJ...264..337S}. We comment on the 
magnitude system and the Johnson-Cousins system in particular
(\S~\ref{mag:sec}).  We find that any ambiguity can be removed with
the proper definition of the \kcorr\ for which we derive the
equations for both photon and energy systems (\S~\ref{kcorr:sec}).  We
conclude that although the differences between the two \kcorrs\
are small, the distinction between energy and photon systems is
important for planned future high-precision supernova experiments
(\S~\ref{con:sec}).

\subsection{Magnitude Systems}
\label{mag:sec}

The primary standards of a photometric system can have their
magnitudes measured either by their energy or photon flux ratios.
Unless a photon--energy conversion correction is later applied, the
flux system is determined by the detectors used to measure the
primaries.  The type of detector used in subsequent observations does
not determine whether the magnitude system is photon or energy based;
in principle the color and airmass corrections put observed magnitudes
into the primary system.

The Johnson-Cousins magnitude system prevalent today is a photon
system, what \citet{Johnson:1953} describe as ``a system of
photoelectric photometry''.  As described in
\citet{Johnson:1951}, their observational setup employed
a photomultiplier as a detector, with the counts being the number of
``deflections'' recorded by a potentiometer.  After an airmass
correction these counts were directly converted to magnitudes. The
secondary stars of \citet{la:1973,la:1983,la:1992} (whose raw data
also were obtained with photon counters) are calibrated via Johnson
and Cousins primary standards and thus must be in the photon system.
Observed magnitudes are therefore photon-based and should be analyzed
as such.

An illustrative example of where there is a numerical difference
between the two magnitude systems is a star that has the same
integrated $B$-band energy flux as Vega (which for simplicity we
consider to be the zero point of the magnitude system) but has a
different photon flux since it has a different spectral energy
distribution (SED). Relative magnitude measurements with a single
filter of a set of stars with similar spectral energy distributions
are independent of whether we are photon counting or measuring energy;
two stars with the same SED but differing brightness will have
$$\Delta m = m_2^\gamma - m_1^\gamma = m_2^\epsilon - m_1^\epsilon
\nonumber$$ where $m_1$ and $m_2$ are the stars' magnitudes.  It
follows that since the zeropoint of magnitude system is based on Vega,
the energy and photon magnitudes of A0V stars are identical:
$m_{A0V}^\gamma = m_{A0V}^\epsilon$.

As an aside, one of the \citet{Johnson:1953} criteria for a
photometric system is ``a determination of the zero point of the color
indices in terms of a certain kind of star which can be accurately
defined spectroscopically.''  Such knowledge, along with the shapes of
the pass-band transmission functions, do allow for calculated
transformation between photon and energy magnitude systems.  Indeed,
much effort has been placed in measuring and modeling the intrinsic
SED of Vega (\citet{dr:1980} and references therein).

\subsection{The \kcorr}
\label{kcorr:sec}

We explicitly review the \kcorr\ calculation of
\citet{kim_kcorr96} that has been used in SCP cosmological analysis. 
To remove any ambiguity
we define the \kcorr\ $K_{xy}$ such that
\begin{equation}
m_y^\alpha=M_x^\alpha + \mu(z) +K_{xy}^\alpha
\label{definition}
\end{equation}
where $\alpha=\{\gamma,\epsilon\}$ for photon or energy magnitude
systems.  The observed magnitude in passband $y$ is $m_y$ and the
absolute magnitude in passband $x$ is $M_x$.  We adopt the theoretical
expression for the distance modulus, $\mu$, based on luminosity
distance.  In other words, the functional form of
$\mu(z;H_0,\Omega_M,\Omega_\Lambda)$ in Equation~\ref{definition} is
identical for photon and energy systems.  Given $f_\lambda(\lambda)$
as the energy flux density of a supernova 10 parsecs away, we can
compute the corresponding energy and photon fluxes at high redshift.
\[
\begin{array}{lr}
f_\lambda(\lambda)d\lambda  & \mbox{Energy flux density in d$\lambda$ bin of a supernova 10
parsecs away}\\
n_\lambda(\lambda)d\lambda = \frac{\lambda d\lambda}{hc}f_\lambda(\lambda) & \mbox{Photon flux in d$\lambda$ bin of a supernova 10 parsecs away}\\
f^z_\lambda(\lambda)d\lambda  =  \frac{d\lambda}{1+z}f_\lambda\left(\frac{\lambda}{1+z}\right)\left(\frac{10 pc}{d_L(z)}\right)^2 & \mbox{Energy flux density in d$\lambda$ bin of a supernova at $z$}\\
n^z_\lambda(\lambda)d\lambda  =  \frac{\lambda d\lambda}{hc(1+z)}f_\lambda\left(\frac{\lambda}
{1+z}\right)\left(\frac{10 pc}{d_L(z)}\right)^2 & \mbox{Photon flux density in d$\lambda$ bin of a supernova at $z$}
\end{array}
\]
The $(1+z)^{-1}$ terms in the redshifted flux densities are due to
wavelength dilution \citep{ok:kcorr}.
The ratio between high and low-redshift photon flux is a factor
$1+z$ greater than the corresponding ratio for energy flux which suffers
from redshifted energy loss.  More precisely
\begin{equation}
\frac{n^z_\lambda(\lambda)}{n_\lambda(\lambda/(1+z))}=\frac{(1+z)f^z_\lambda(\lambda)}{f_\lambda(\lambda/(1+z))}.
\label{ratio}
\end{equation}
The fact that the relative photon fluxes of high-redshift supernovae are
$1+z$ ``brighter'' than energy fluxes can be interpreted as being due
to the latter's extra energy loss due to redshift. 

Using the fact that $\mu=-5\log{\left(\frac{10 pc}{d_L(z)}\right)}$ we can compute and
compare energy and photon \kcorrs,
\begin{eqnarray}
  K_{xy}^\epsilon & = &  -2.5 \log
    \left(
    \frac
       {\int {\cal Z}^\epsilon_x(\lambda)S_x(\lambda)d\lambda}
       {\int {\cal Z}^\epsilon_y(\lambda)S_y(\lambda)d\lambda}
    \right)
    +2.5 \log(1+z) 
    +2.5 \log
    \left( 
    \frac
	{\int f_\lambda(\lambda)S_x(\lambda)d\lambda}
	{\int f_\lambda(\lambda/(1+z))S_y(\lambda)d\lambda}
    \right)
\label{ekcorr}
\end{eqnarray}
\begin{eqnarray}
  K_{xy}^{\gamma} & = & -2.5 \log
    \left(
    \frac
       {\int \lambda{\cal Z}^\gamma_x(\lambda)S_x(\lambda)d\lambda}
       {\int \lambda{\cal Z}^\gamma_y(\lambda)S_y(\lambda)d\lambda}
    \right)
    +2.5 \log(1+z) 
    +2.5 \log
    \left( 
    \frac
	{\int \lambda f_\lambda(\lambda)S_x(\lambda)d\lambda}
	{\int \lambda f_\lambda(\lambda/(1+z))S_y(\lambda)d\lambda}
    \right).
\label{gkcorr}
\end{eqnarray}
The filter transmission functions are given as $S_i(\lambda)$ where
$S_x$ is the rest-frame filter and $S_y$ is the observer filter.  (The
transmission functions give the fraction of photons transmitted at a
given wavelength where we assume no down-scattering.)  For the
standard star (i.e. calibrator) SED ${\cal Z}(\lambda)$ we assume the
existence of a standard star with identical properties as the
supernova, i.e. with exactly the same color and observed through the
same airmass.  Pragmatically, this assumption affirms perfect
photometric calibration to all orders of color and airmass.  For
convenience, we choose these secondary standards to have 0 magnitude.
In principle, a different standard will be needed for each filter,
choice of photon or energy flux, and each source SED.  Each standard
is labeled ${\cal Z}_X^\alpha$ where $X=\{U,B,V,R,I,\ldots\}$ and
$\alpha=\{\gamma,\epsilon\}$ for photon or energy flux as defined
earlier.

Equations~\ref{ekcorr} and \ref{gkcorr} generalize the \kcorrs\
of \citet{1983ApJ...264..337S}\footnote{Note that in the notation of
\citet{1983ApJ...264..337S}, $f_\nu$ and $f_{\nu(1+z)}$ are the same
function evaluated at different frequencies.}  and are precisely those
given and calculated in KGP96.  In that paper, it was
found that the differences between the two \kcorrs\ are non-zero
but small, $|K_{xy}^\epsilon-K_{xy}^{\gamma}|<0.07$ magnitudes.  They
are a function of redshift, filters, and supernova epoch and thus can
cause small systematic shifts in light-curve shapes and magnitude
deviations in the Hubble diagram.  The use of the incorrect
\kcorr\ will have a significant effect on experiments with small
$< 0.1$ targeted magnitude errors.

To illustrate, in Figure~\ref{counts} we plot $K_{BZ}^\epsilon -
K_{BZ}^\gamma$ (where $Z$ refers to the passband and not redshift) for
a standard Type Ia supernova at $B$ maximum and 15 rest-frame days
after maximum out to $z=2$.  The differences are close to zero at $z
\sim 1.1$ where $B(\lambda/(1+z)) \sim Z(\lambda)$.  Beyond this
optimal redshift, the differences can be $>0.01$ magnitudes.  The
redder color of the supernova at the later epoch gives relatively larger
photon \kcorrs\ over almost all redshifts.

The similarity in the two \kcorrs\ is due to two competing terms
that nearly cancel.  A photon \kcorr\ is $1+z$ brighter because
the supernova does not suffer redshifting energy loss.  However, the
zeropoint of the redder filter used to observe the redshifted
supernova is larger since an A0V photon spectrum is flatter than its
energy spectrum.  This makes the observed supernova magnitude
numerically fainter.  Consider the special case where
$S_y(\lambda)=S_x(\lambda/(1+z))$.  With perfect filter-matching the
specifics of the supernova spectrum are unimportant and the
\kcorrs\ depend on the zeropoints:
\begin{eqnarray}
  K_{xy}^\epsilon & = &  -2.5 \log
    \left(
    \frac
       {\int {\cal Z}^\epsilon_x(\lambda)S_x(\lambda)d\lambda}
       {\int {\cal Z}^\epsilon_y(\lambda)S_y(\lambda)d\lambda}
    \right)\\
  K_{xy}^{\gamma} & = & -2.5 \log
    \left(
    \frac
       {(1+z)\int \lambda{\cal Z}^\gamma_x(\lambda)S_x(\lambda)d\lambda}
       {\int \lambda{\cal Z}^\gamma_y(\lambda)S_y(\lambda)d\lambda}
    \right)\\
 &=& -2.5 \log
    \left(
    \frac
       {(1+z) <\lambda_x>\int {\cal Z}^\gamma_x(\lambda)S_x(\lambda)d\lambda}
       {<\lambda_y>\int {\cal Z}^\gamma_y(\lambda)S_y(\lambda)d\lambda}
	\right)
\end{eqnarray}
where $<\lambda>$ is the effective wavelength of the standard through
the filter.  As long as the standard star is well behaved, we expect
the effective wavelength of the redshifted filter to be $1+z$ greater
then that of the restframe filter $<\lambda_y> \sim (1+z)<\lambda_x>$
so that
\begin{equation}
	K_{xy}^{\gamma} \sim -2.5 \log
	\left(
    \frac
       {\int {\cal Z}^\epsilon_x(\lambda)S_x(\lambda)d\lambda}
       {\int {\cal Z}^\epsilon_y(\lambda)S_y(\lambda)d\lambda}
    \right) = K_{xy}^{\epsilon}.
\end{equation}
Choosing filters that accept the same spectral region at both low and
high redshifts not only reduces errors but also reduces the difference
between energy and photon \kcorrs.

The effect of using the ``energy'' distance modulus in defining the
\kcorr\ in Equations~\ref{definition}, \ref{ekcorr}, and
\ref{gkcorr} are seen in the open-filter \kcorrs.  When
$S_x=S_y=1$, the energy \kcorr\ is unnecessary and indeed
$K_{xy}^\epsilon=0$. For the photon \kcorr\ we find
$K_{xy}^\gamma=-2.5\log{(1+z)}$, the difference between ``energy'' and
``photon'' distance moduli.

A simple measure for the difference between single-filter
\kcorrs\ is the ratio in effective wavelength of a redshifted and
unredshifted source through that filter.  For example, sources with
power-law SED's have identical photon and energy \kcorrs.  For
low-redshift objects the difference in effective wavelength should be
very small (unless they have pathological spectra) and thus make
little difference in distance determinations.  For example, a Type Ia
supernova at maximum at $z=0.1$ observed through the $B$-band would
have a distance modulus error of 0.02 magnitudes if the wrong
\kcorr\ were applied.

\subsection{Final Notes on \kcorrs}
\label{con:sec}
In this appendix we have shown that the measurements $m_Y(z)-M_X$ do
depend on whether the magnitude system is based on energy or photon
flux.  Although the ``photon luminosity distance'' is shorter than the
standard luminosity distance, we can still use the relation
$m_Y(z)=M_X+\mu(z)+K_{XY}$ with the appropriate definitions of the
\kcorrs; the ones of
\citet{kim_kcorr96} are appropriate.  With this definition, the
standard equations linking the energy distance modulus to cosmology
are applicable.  The Johnson-Cousins magnitude system is in fact
photon-based.  Therefore, the $K^\gamma_{XY}$ \kcorr\ should and
has been used in the supernova cosmology analysis of the SCP. 
Although application of the incorrect \kcorr\
would contribute negligibly to the error budget of the current
supernova sample, the distinction is important for precision
experiments that require 0.02 magnitude accuracies, such as the
Supernova Acceleration Probe.  With the choice of well-matched
filters, differences between energy and photon \kcorr\ can be
minimal.

Using the ``count'' distance modulus based on $d_\gamma$ in
Equation~\ref{definition} would provide a more physically satisfying
definition of the count \kcorr.  Recall that $\mu^\epsilon=
\mu^\gamma + 2.5\log{(1+z)}$. Then the extra $2.5\log{(1+z)}$ in the
\kcorr\ would give
\begin{equation}
  K_{xy}^{\gamma} = -2.5 \log
    \left(
    \frac
       {\int \lambda{\cal Z}^\gamma_x(\lambda)S_x(\lambda)d\lambda}
       {\int \lambda{\cal Z}^\gamma_y(\lambda)S_y(\lambda)d\lambda}
    \right) 
    +2.5 \log
    \left( 
    \frac
	{\int \lambda f_\lambda(\lambda)S_x(\lambda)d\lambda}
	{\int \lambda' f_\lambda(\lambda')S_y((1+z)\lambda')d\lambda'}
    \right).
\label{gkcorr2}
\end{equation}
In other words, the \kcorr\ would depend simply on the ratio of
supernova photons in the rest-frame filter and a blue-shifted observer
filter, and the zeropoint.  This methodology would preserve the
physical meanings that we associate with both distance modulus and
\kcorr.  For simplicity, however, we here adopt the energy
distance modulus for both \kcorrs\ to be consistent with the
literature and to ensure unambiguity when referring to $K$-corrected
magnitudes and distance moduli.

\clearpage

\begin{deluxetable}{cccccc}
\tablewidth{0pt}
\tablecaption{List of SNe~Ia used to create the spectral template}
\tablehead{\colhead{SN Name} &\colhead {Epochs with respect to maximum light}
&\colhead{References}} 
\startdata
1981B &   0,17,20,26,35,49,58,64 &  1 \\
1989B &  -7,-5,-3,-2,-1,3,5,8,9,11,12,13,14,15,16,17,18,19 & 2  \\
1990N &  -14,-7,7,14,17,38 &  3,4,5 \\
1992A &  -5,-1,3,5,6,7,9,11,16,17,24,28,37,46,76 & 6 \\
1994D &  -10,-9,-8,-7,-4,-3,2,3,4,5,6,8,11,12,13,14,16,18,20,25 & 7,8 \\
1990af & -2  & 9 \\ 
1992ag &  0 &  9 \\ 
1992al & -5, 4 & 9 \\ 
1992aq &  2 & 9 \\ 
1992bc & -10 &  9 \\ 
1992bh &  3 &  9 \\ 
1992bl &  2 &  9 \\ 
1992bo &  2 & 9 \\ 
1992bp &  7 & 9 \\ 
1992br &  7 & 9 \\ 
1992bs &  3 & 9 \\ 
1992P  &  0 & 9 \\ 
1993H  &  3, 5 & 9 \\ 
1993O  & -4, 2 & 9 \\ 
\enddata
\tablerefs{(1) \citet{bra81b}, (2) \citet{wells89b},
(3) \citet{leibun1}, (4) \citet{phil91t}, (5) \citet{maz90n}, (6)
\citet{kir92a} (7) \citet{meik94d}, (8) \citet{patat94d}, (9) Courtesy of
the Cal\'an/Tololo Supernova Survey. Note that, where applicable, the
{\it IUE} spectra from \citet{iuesne} were used in forming the template.} 
\label{templ_tab}
\end{deluxetable}

\clearpage

\begin{deluxetable}{ccccccccc}
\tablewidth{0pt}
\tablecaption{$UBVRIJHK$ magnitude differences from the peak $B$-band
value for the spectral template}
\tablehead{\colhead{Time} & \colhead {U} & \colhead{B} & \colhead{V}
& \colhead{R} &\colhead{I} & \colhead{J} & \colhead{H} & \colhead{K}} 
\startdata
-20.0 & $\infty$ & $\infty$ & $\infty$ & $\infty$ & $\infty$ &
$\infty$ & $\infty$ &$\infty$\\ 
-15.0 &  2.16 & 2.37 & 2.71 & 2.66 & 3.11 & 3.27 & 3.45 & 3.28 \\
-10.0 &  0.43 & 0.83 & 1.13 & 0.84 & 1.11 & 1.69 & 1.87 & 1.70 \\
 -5.0 & -0.25 & 0.13 & 0.26 & 0.21 & 0.48 & 0.82 & 1.00 & 0.83 \\
  0.0 & -0.22 & 0.00 & 0.06 & 0.07 & 0.42 & 0.62 & 0.80 & 0.63 \\
  5.0 &  0.13 & 0.16 & 0.08 & 0.05 & 0.53 & 0.72 & 0.90 & 0.73 \\
 10.0 &  0.75 & 0.55 & 0.27 & 0.41 & 0.87 & 0.82 & 1.00 & 0.83 \\
 15.0 &  1.45 & 1.05 & 0.53 & 0.67 & 1.00 & 1.86 & 1.13 & 1.14 \\
 20.0 &  2.15 & 1.59 & 0.76 & 0.66 & 0.77 & 2.26 & 0.93 & 0.90 \\
 25.0 &  2.65 & 2.08 & 1.03 & 0.81 & 0.72 & 1.94 & 0.73 & 0.73 \\
 30.0 &  2.91 & 2.49 & 1.50 & 1.09 & 0.87 & 1.52 & 0.57 & 0.68 \\
 35.0 &  3.03 & 2.78 & 1.85 & 1.42 & 1.30 & 1.64 & 0.88 & 1.00 \\
 40.0 &  3.15 & 2.99 & 2.12 & 1.67 & 1.63 & 1.93 & 1.17 & 1.33 \\
 45.0 &  3.26 & 3.15 & 2.34 & 1.88 & 1.88 & 2.42 & 1.46 & 1.60 \\
 50.0 &  3.37 & 3.25 & 2.50 & 2.02 & 2.03 & 2.91 & 1.75 & 1.88 \\
 55.0 &  3.45 & 3.33 & 2.63 & 2.19 & 2.23 & 3.21 & 1.95 & 2.08 \\
 60.0 &  3.53 & 3.41 & 2.77 & 2.37 & 2.44 & 3.51 & 2.15 & 2.28 \\
 65.0 &  3.62 & 3.50 & 2.92 & 2.45 & 2.53 & 3.81 & 2.33 & 2.46 \\
 70.0 &  3.70 & 3.58 & 3.06 & 2.54 & 2.61 & 4.12 & 2.52 & 2.64 \\
\enddata
\label{stretch_col}
\end{deluxetable}

\clearpage

\begin{deluxetable}{ccccccc}
\tablewidth{0pt}
\tablecaption{$K_{BB}(z=0.1)$ for two SNe~Ia}
\tablehead{
\colhead{} & \multicolumn{3}{c}{SN~1994D} & \multicolumn{3}{c}{SN~1992A}\\
\colhead{$t - t_0$} & \colhead {$U-B$} & \colhead {$B-V$}
& \colhead{$K_{BB}(z=0.1)$} & \colhead {$U-B$} & \colhead {$B-V$} &
\colhead{$K_{BB}(z=0.1)$}}
\startdata
 -5 & -0.6 & -0.07 & -0.04 & -0.3  & -0.03 & 0.09 \\
  0 & -0.5 & -0.08 & -0.02 & -0.3  & -0.01 & 0.07 \\
  5 & -0.3 &  0.04 &  0.06 & -0.1  &  0.26 & 0.15 \\
 10 & -0.2 &  0.22 &  0.10 &  0.1  &  0.40 & 0.24 \\
 15 &  0.0 &  0.56 &  0.20 &  0.3  &  0.74 & 0.34 \\
 20 &  0.2 &  0.90 &  0.33 &  0.4  &  1.11 & 0.42 \\
\enddata
\label{ulow}
\end{deluxetable}

\clearpage

\setcounter{figure}{0}

\begin{figure}[p]
\psfig{file=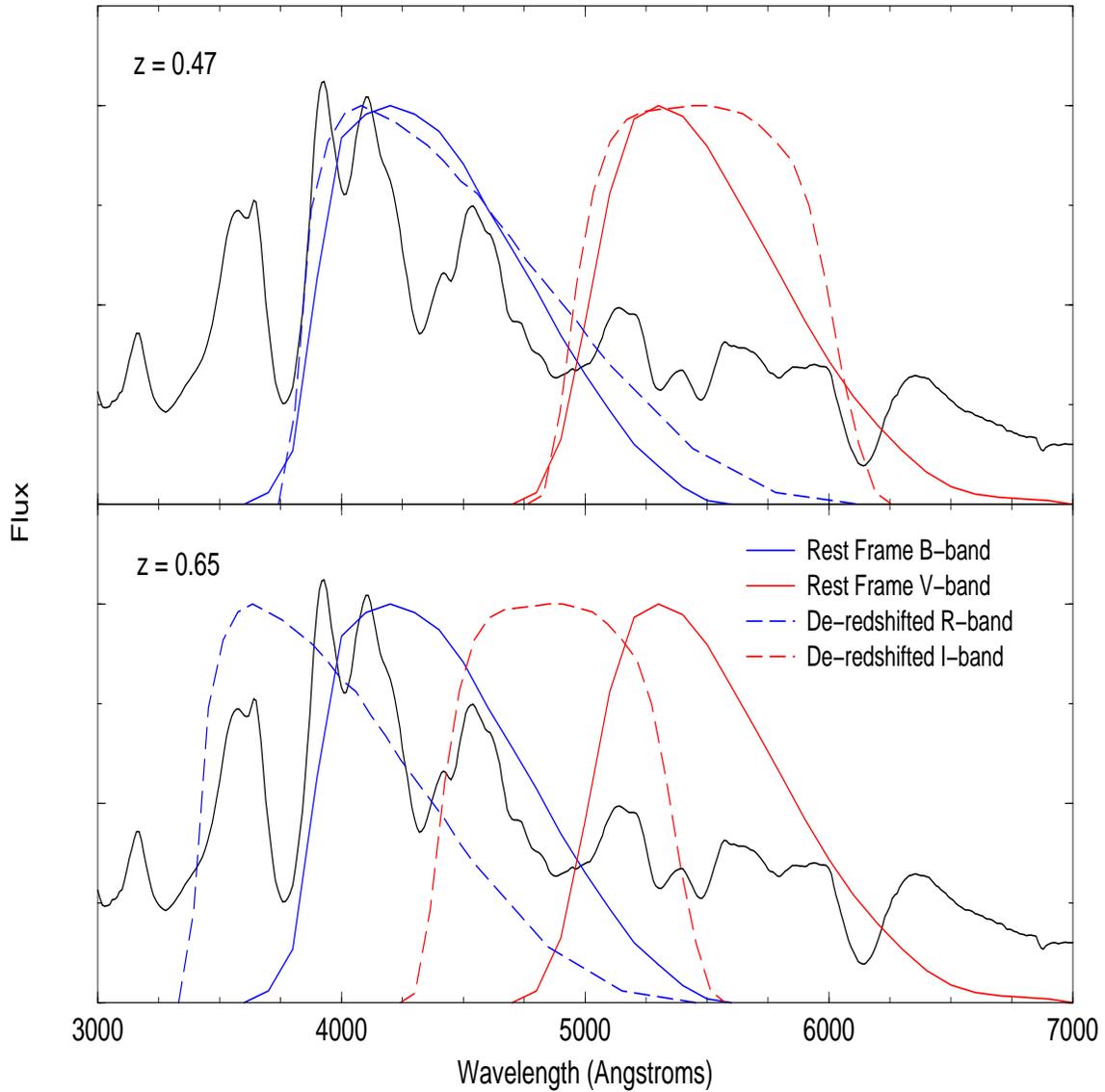,height=6.0in,width=6.0in,angle=270}
\caption{Filter pairings for $R$- and $I$-band observations of
a redshifted SN~Ia near maximum-light with respect to rest-frame $B$-
and $V$-band filters. Nice matches occur at $z=0.47$ while large
extrapolations occur at $z=0.65$.\label{kcorr_flt}}
\end{figure}

\clearpage

\begin{figure}[p]
\psfig{file=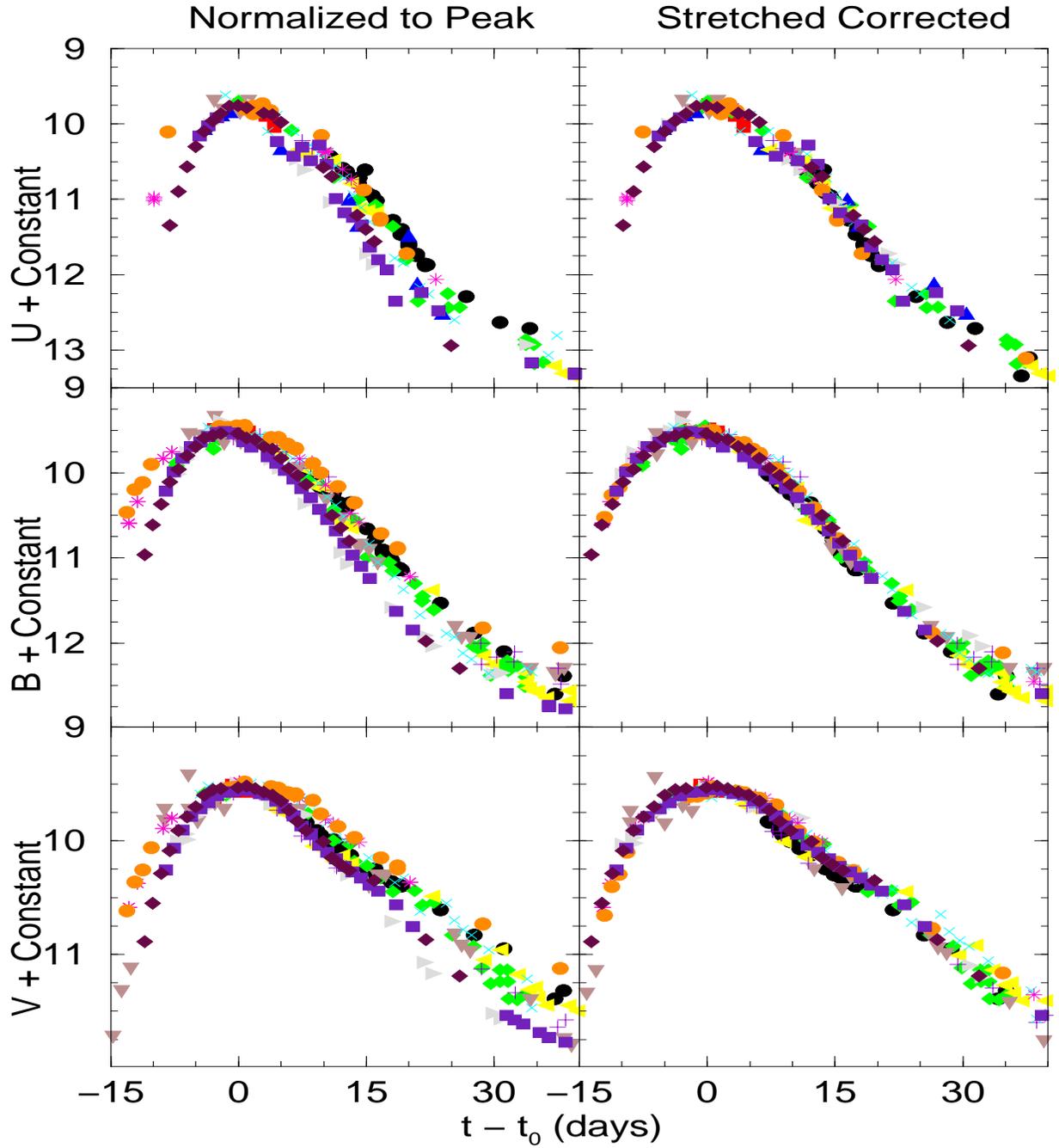,height=7.0in,width=7.0in,angle=270}
\caption{{\it Left panels}: The $UBV$-band light curves of several
nearby SNe~Ia adjusted to the same peak magnitude. {\it Right panels}:
The same light curves {\it corrected} by their $B$-band stretch
factor. Note the large reduction in scatter about the
mean.\label{phot_all}}
\end{figure}

\clearpage

\begin{figure}[p]
\psfig{file=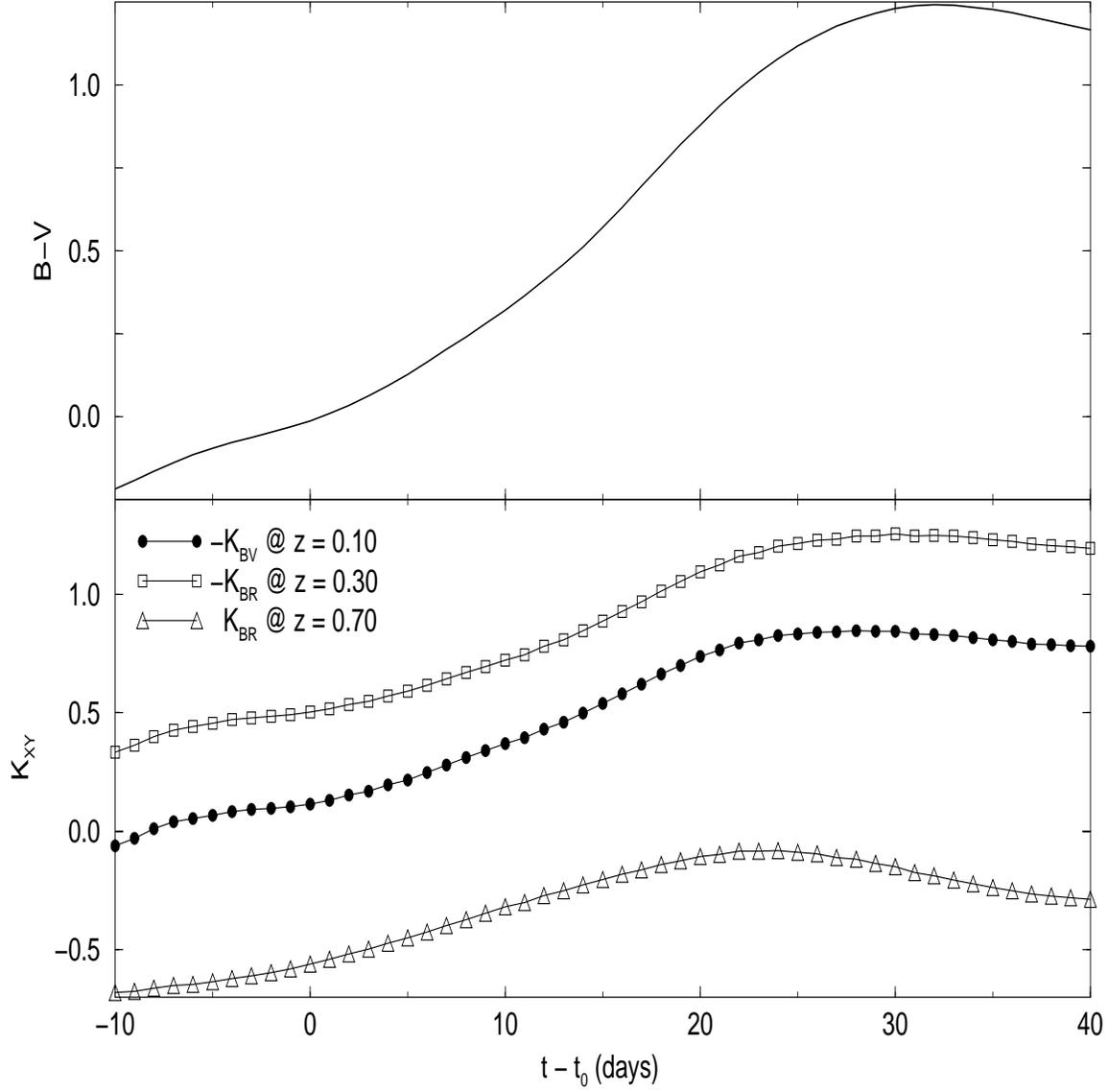,height=6.0in,width=6.0in,angle=270}
\caption{{\it Top panel}: Rest-frame $B-V$ color curve for a $s=1.0$
SN~Ia. {\it Bottom panel}: \kcorrs\ for a $s=1.0$ SN~Ia at a variety
of redshifts. Note the strong similarity between the color curve and
the \kcorrs.\label{clrcrv}}
\end{figure}

\clearpage

\begin{figure}[p]
\psfig{file=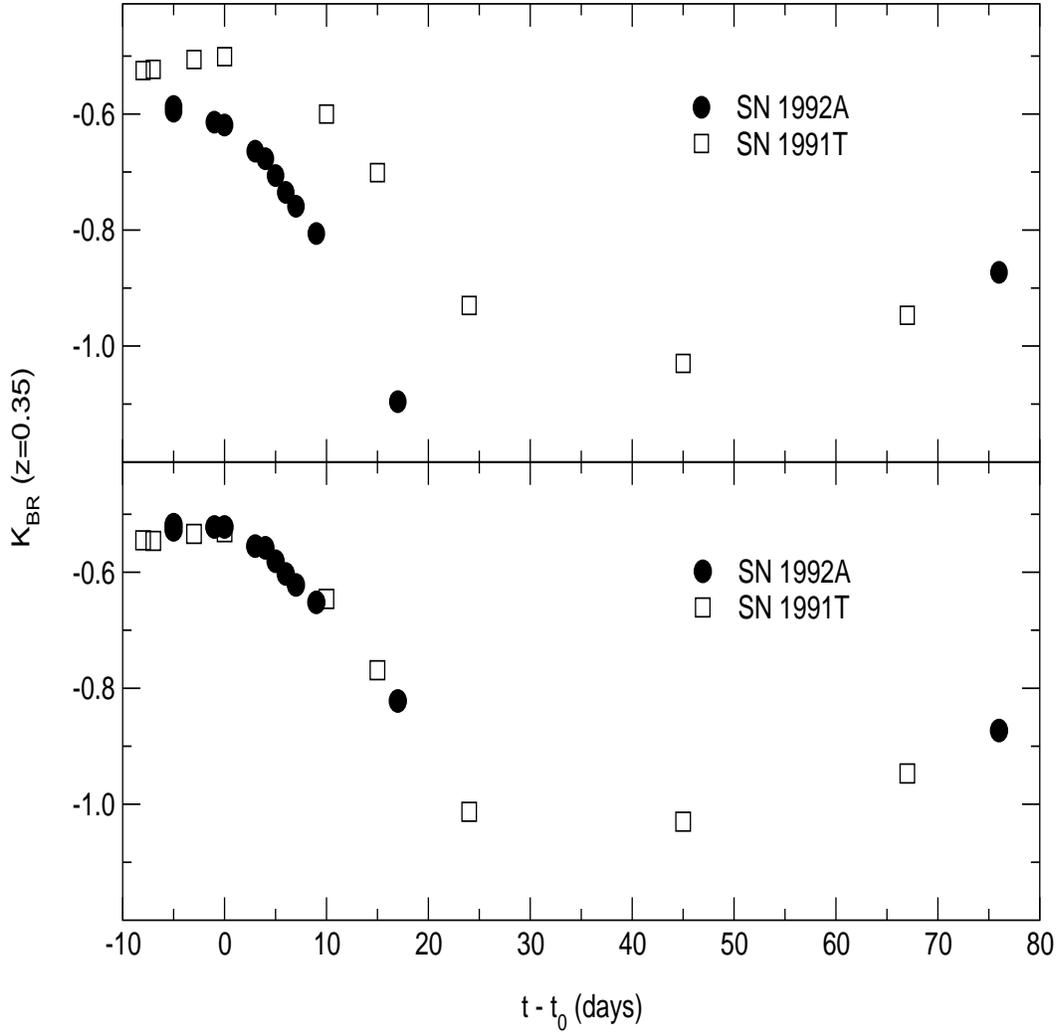,height=6.0in,width=6.0in,angle=270}
\caption{$K_{BR}$ at $z = 0.35$ for SN~1992A (black circles) and
SN~1991T (red squares). {\it Top panel}: \kcorrs\ as would be observed
for these spectroscopically distinct SNe~Ia ($s =$ 0.80 and 1.09
respectively). {\it Bottom panel}: A slope correction was applied to
each of the supernova's spectra so that they would have the same $B-V$
colors as an $s = 1.0$ SN~Ia at every epoch. \label{kc_str}}
\end{figure}

\clearpage

\begin{figure}[p]
\psfig{file=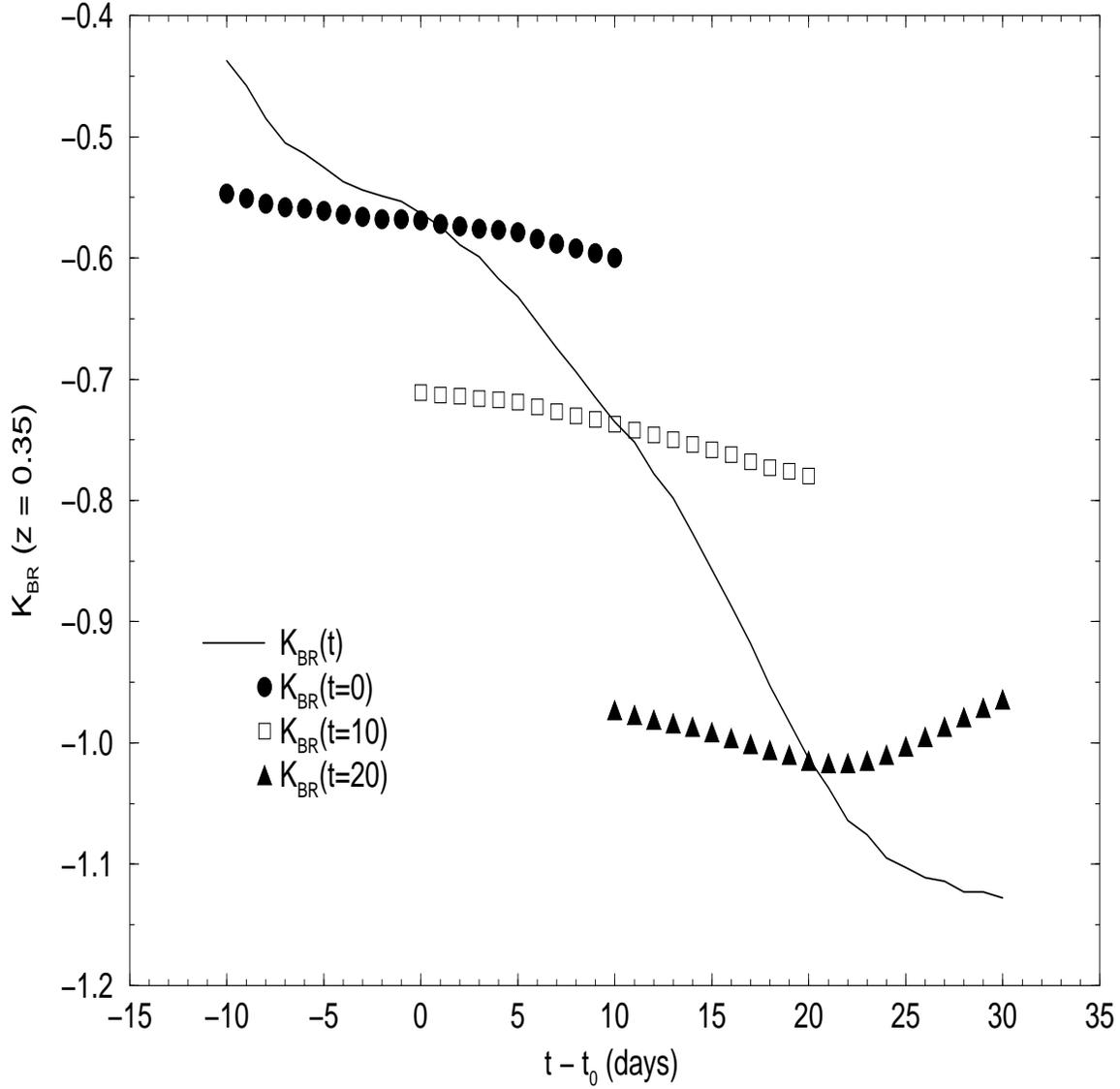,height=6.0in,width=6.0in,angle=270}
\caption{$K_{BR}$ for a $z=0.35$ SN~Ia with $s=1.0$ along with three
other sets of \kcorrs. These \kcorrs\ were calculated by forcing the
spectra within 10 days of a given epoch on the light curve (maximum
light, $+10$ days and $+20$ days) to have the same colors as the SN~Ia
has at that epoch. The difference in the \kcorrs\ is, at most, 0.05 mag
over the entire 20 day window and $\lesssim$ 0.02 mag for a 10 day
window.\label{kc_time}}
\end{figure}

\clearpage

\begin{figure}[p]
\psfig{file=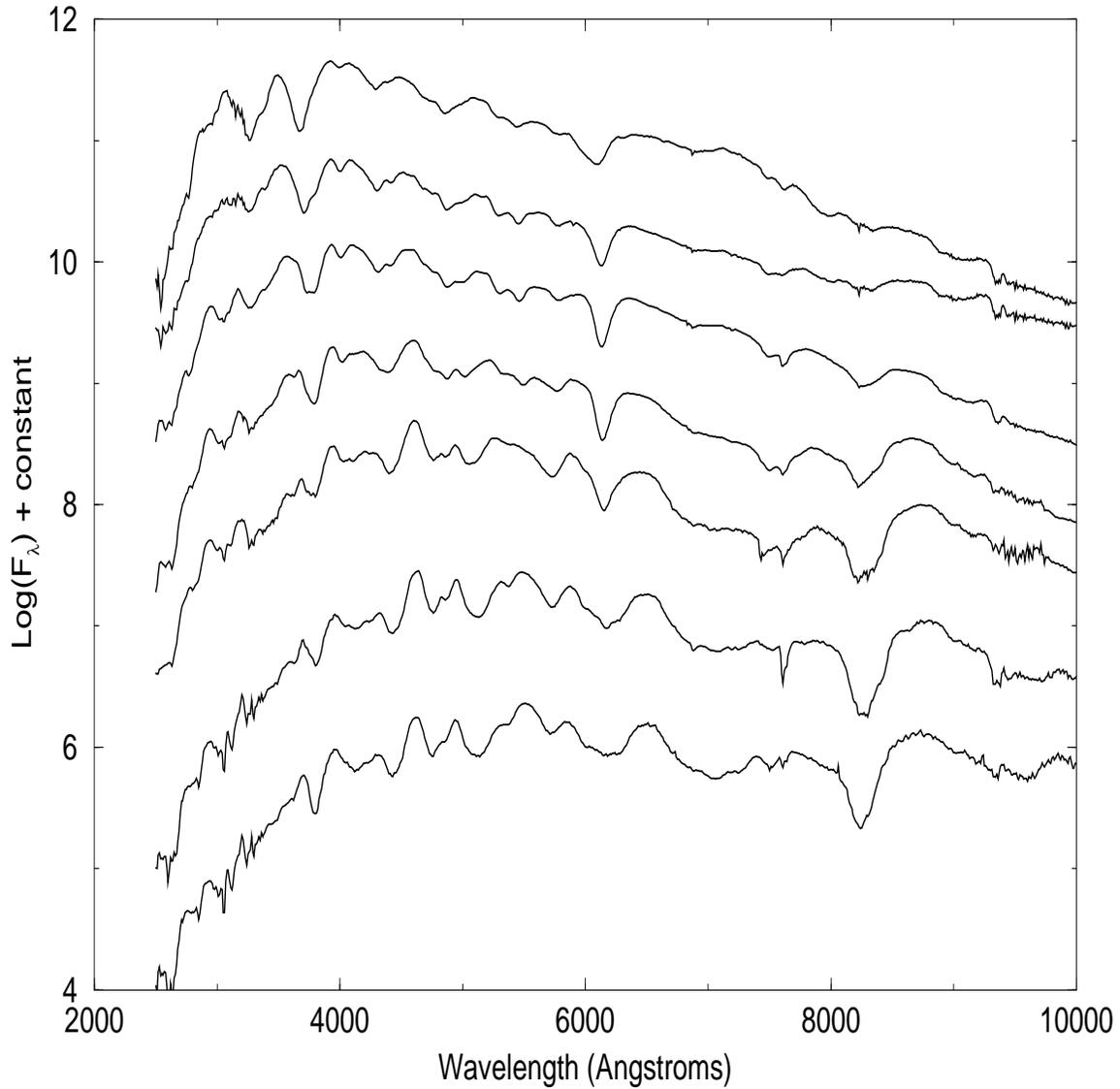,height=6.0in,width=6.0in,angle=270}
\caption{A weekly sampling of the spectral template from two weeks
before to four weeks after maximum light. The spectra and light curves
are available at: {\tt
http://www.supernova.lbl.gov/$\sim$nugent/spectra.html}\label{spectra}} 
\end{figure}

\clearpage

\begin{figure}[p]
\psfig{file=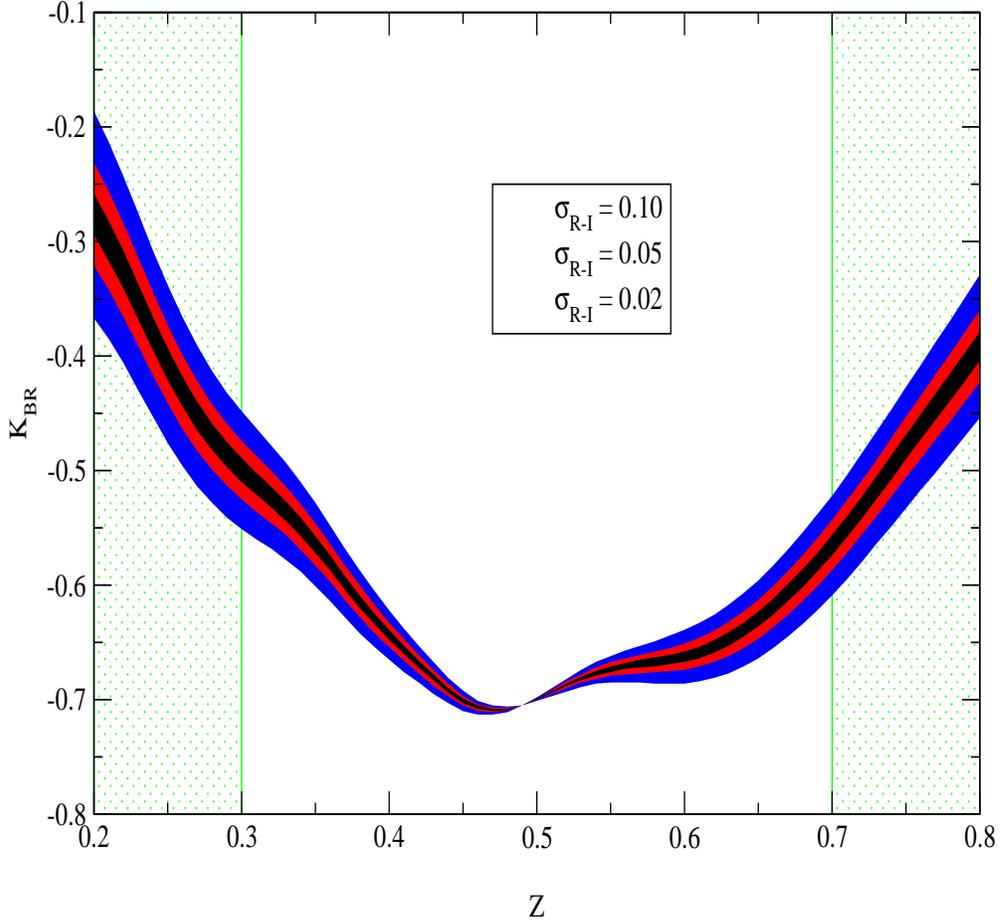,height=6.0in,width=6.0in,angle=270}
\caption{$K_{BR}$ as a function of $z$ for three different cases in
which we assign an uncertainty in the observed $R-I$ color of 0.02
(black), 0.05 (red) and 0.10 (blue). Note how the uncertainty in
$K_{BR}$ vanishes at $z = 0.49$, where the rest-frame $B$ and
de-redshifted $R$-band filters nicely overlap, regardless of the
uncertainty in the color measurement. In practice one does not use
$K_{BR}$ in the green regions, where a switch is made to a more
appropriate filter.\label{kcorr_err}}
\end{figure}

\clearpage

\begin{figure}[p]
\psfig{file=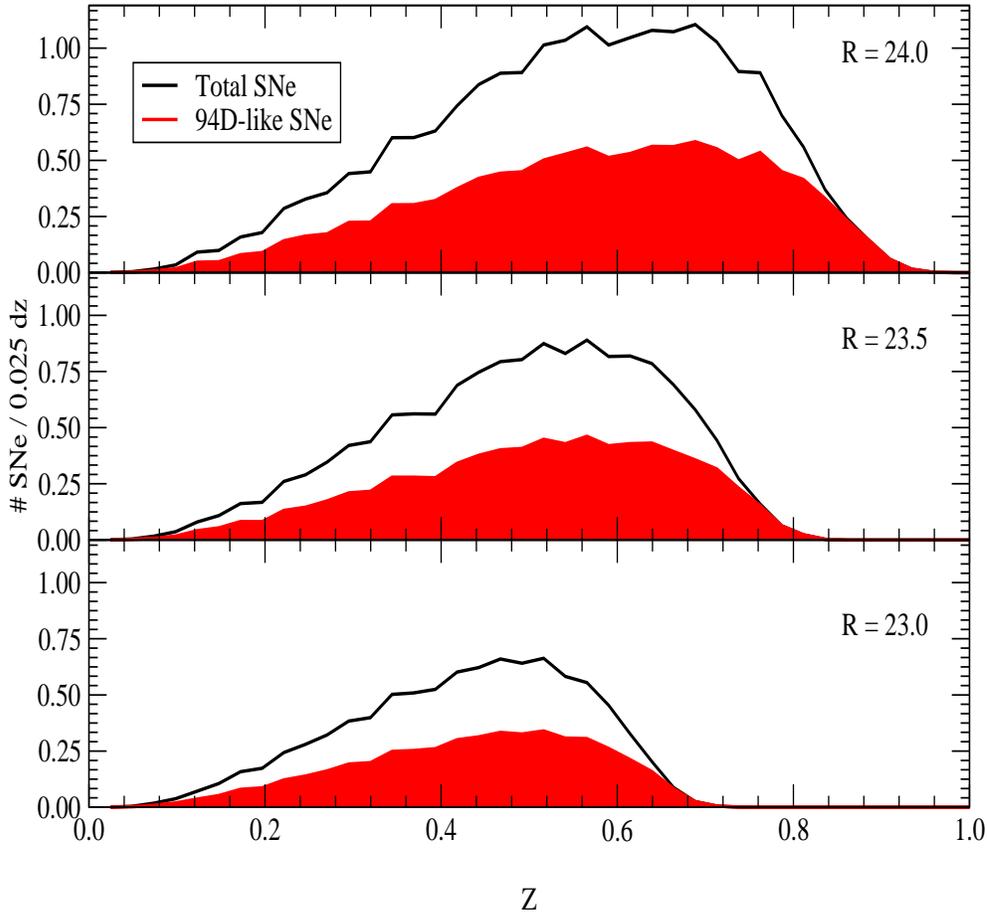,height=6.0in,width=6.0in,angle=270}
\caption{An $R$-band Monte Carlo search for SNe~Ia covering a one
month gap between reference and discovery and 4 square degrees on the
sky given an equal distribution between the blue SN 1994D-like
supernovae and the red SN 1992A-like supernovae. The solid black line
plots the total number of supernovae discovered per 0.025 $dz$ while
the filled red area shows the total fraction which are 1994D-like. The
bias becomes apparent at $z > 0.5$. \label{search_r}}
\end{figure}

\clearpage

\begin{figure}[p]
\psfig{file=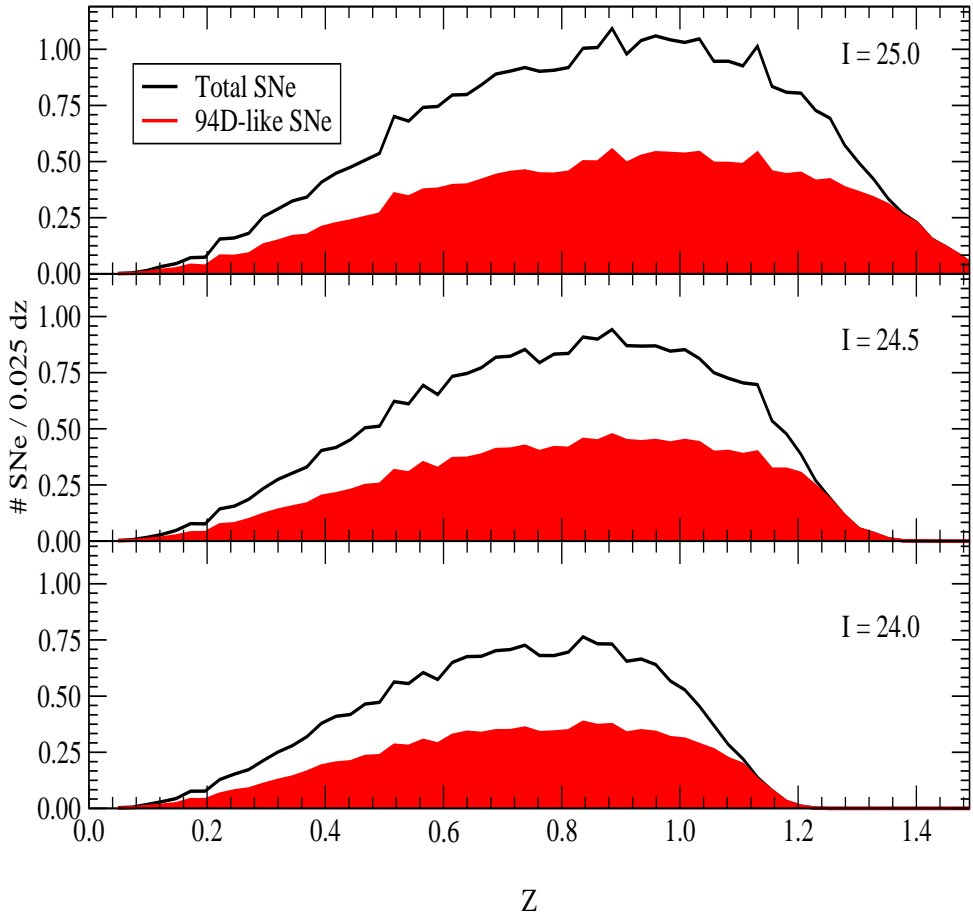,height=6.0in,width=6.0in,angle=270}
\caption{Same as Figure~\ref{search_r} except for an $I$-band search
only covering 1 square degree. Here this bias is moved out to $z >
0.9$ as that is where the restframe $U$-band light enters the $I$-band
filter.\label{search_i}}
\end{figure}

\clearpage

\begin{figure}[p]
\psfig{file=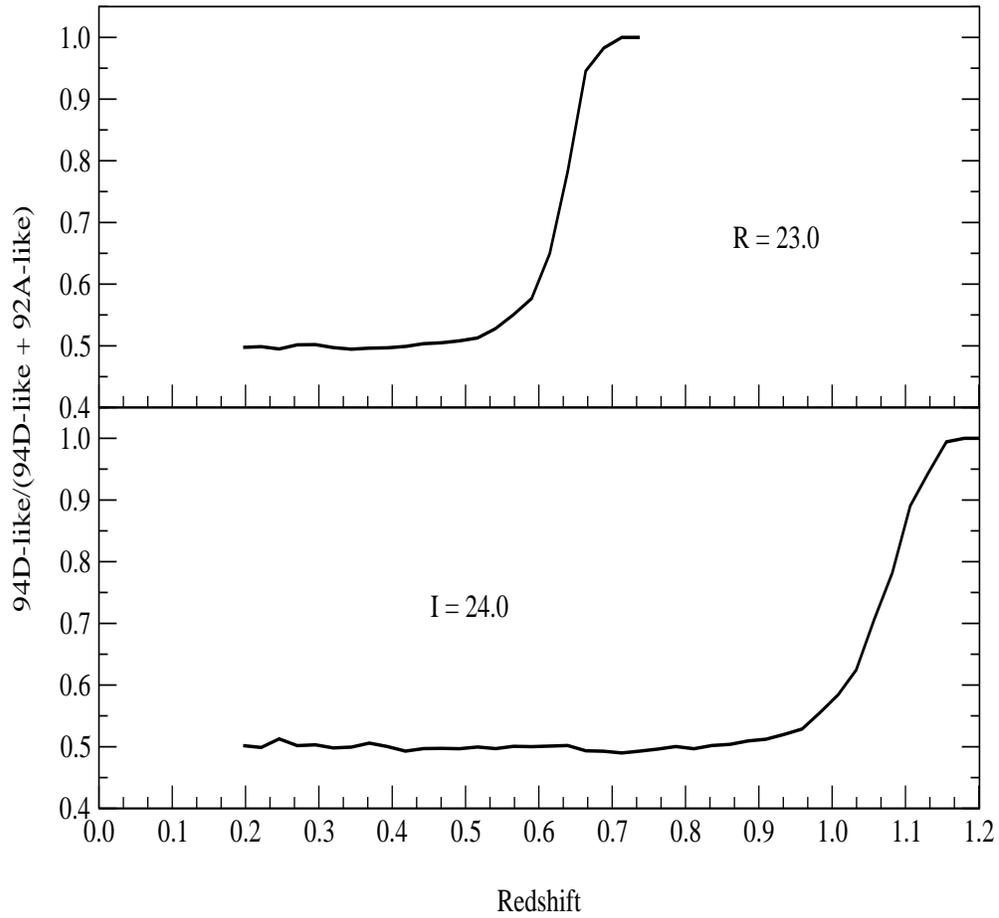,height=6.0in,width=6.0in,angle=270}
\caption{The data from Figures~\ref{search_r}~and~\ref{search_i}
presented as the ratio of SN~1994D-like supernovae to the total number
of supernovae found just the $R$=23.0 and $I$=24.0 limiting magnitude
searches. \label{search_ratio}}
\end{figure}

\clearpage

\begin{figure}[p]
\psfig{file=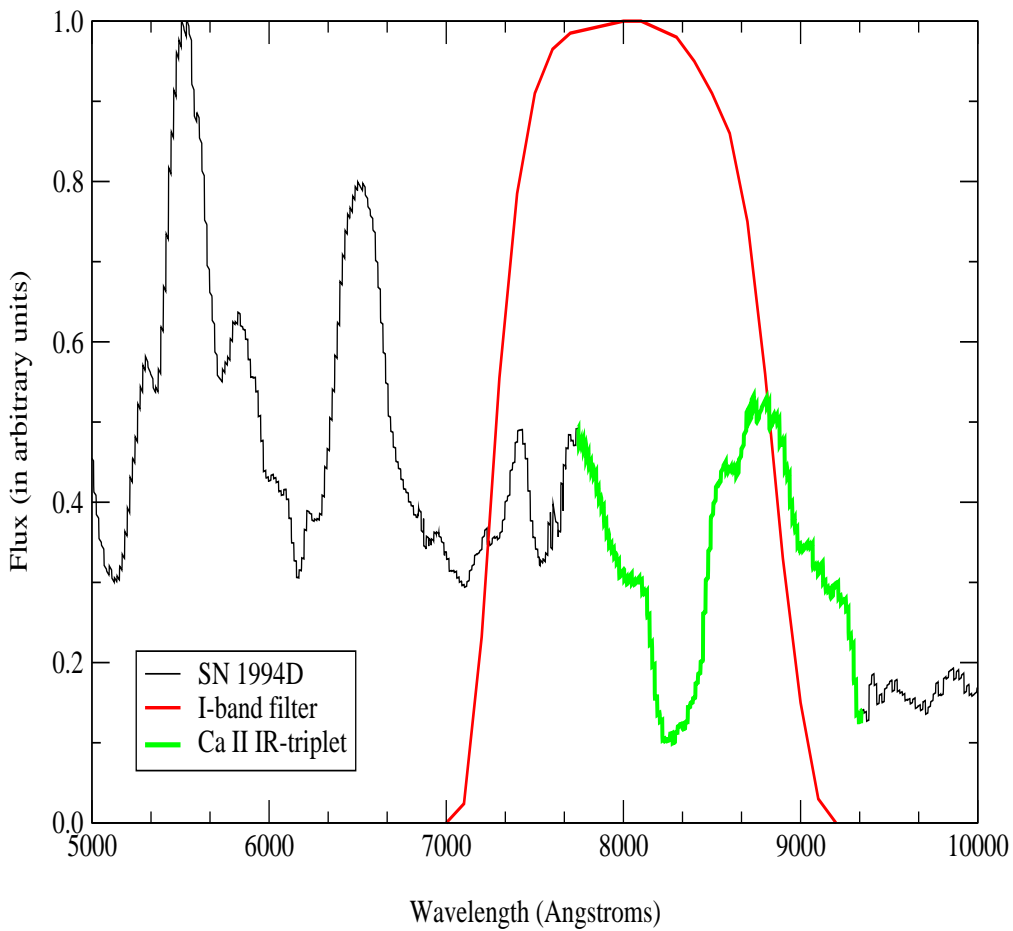,height=6.0in,width=6.0in,angle=270}
\caption{SN 1994D (in black) at 24 days after maximum $B$-band light
\citep{patat94d}. Highlighted are the strong Ca~II IR triplet feature
(in green) and the Bessel $I$-band filter function (in red). Errors in
the \kcorrs\ can be quite significant, even at low redshift, if one
does not compensate for the change in the feature's strength with
epoch and luminosity class.\label{iband}}
\end{figure}

\clearpage

\begin{figure}[p]
\psfig{file=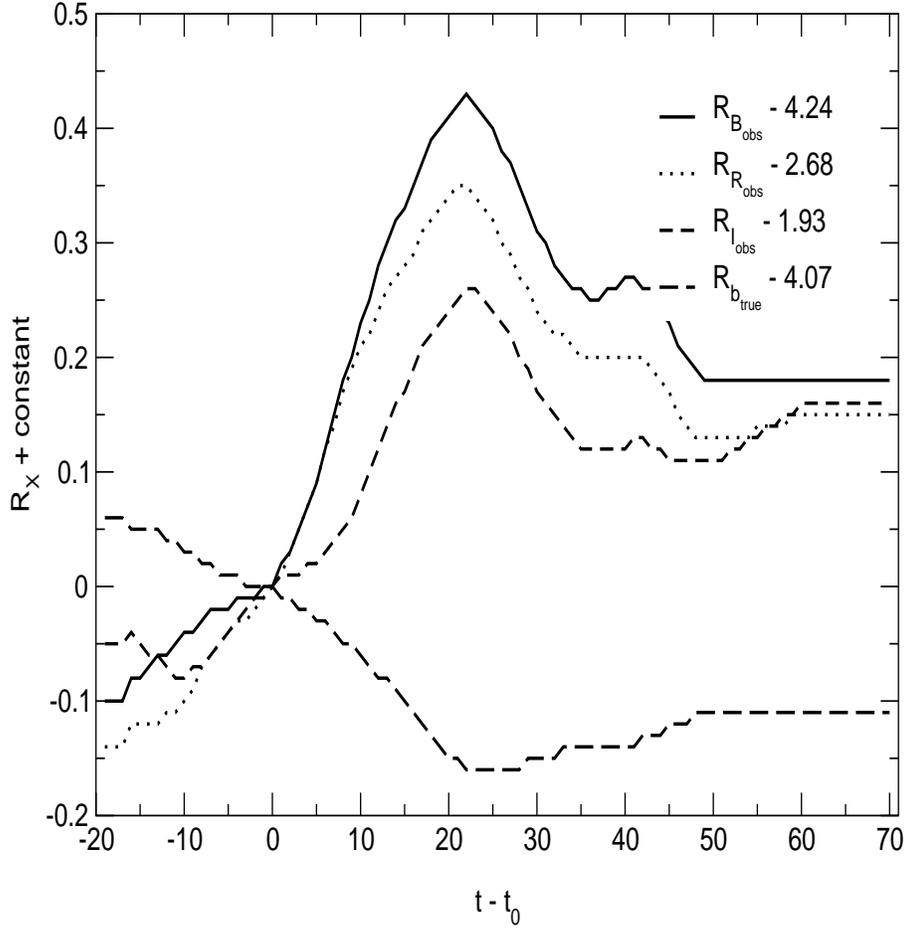,height=6.0in,width=6.0in,angle=270}
\caption{The relative variations of $R_{X} = A_{X}/E(B-V)$ as a
function of time. The calculations were performed for $B$, $R$ and $I$
($V = R_{B} - 1.0$ by definition and thus is not shown). Both
$R_{B_{obs}}$ and $R_{B_{true}}$ are displayed. Each serves a purpose
for performing color corrections depending on whether one wants to
correct photometry based on their observed colors, or if one {\it a
priori} knows $E(B-V)_{true}$ along the line of sight to the SN~Ia in
question.\label{red_time}}
\end{figure}

\clearpage

\begin{figure}[p]
\psfig{file=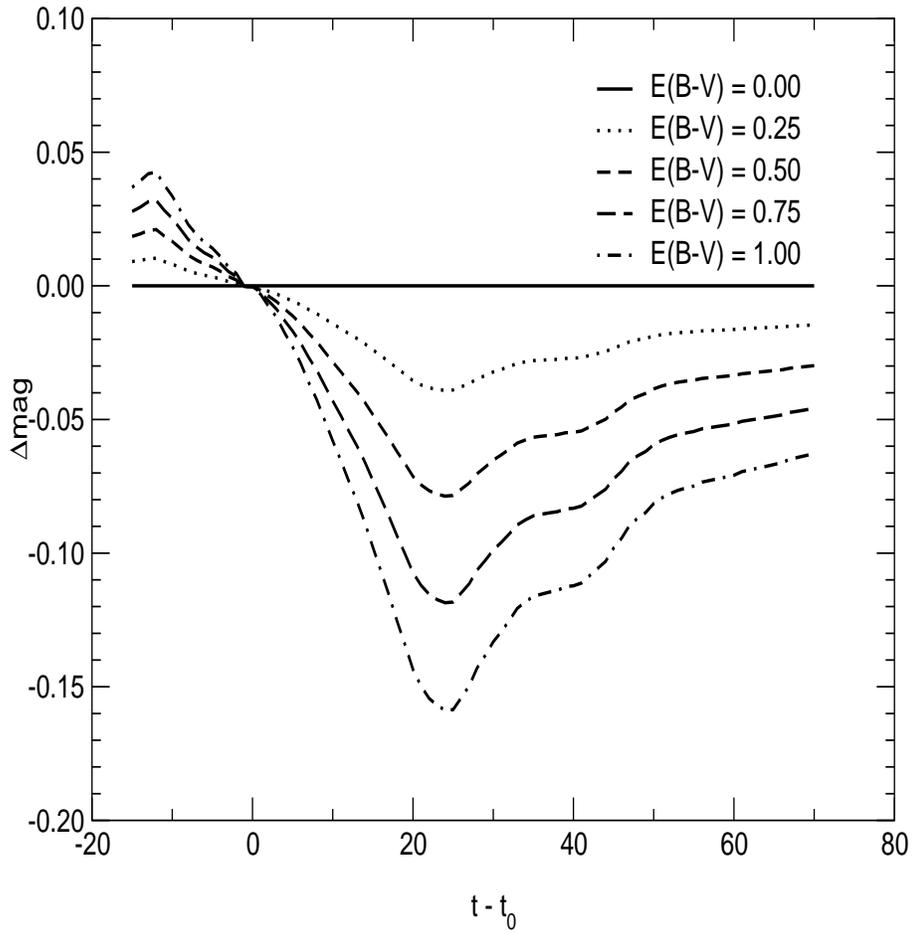,height=6.0in,width=6.0in,angle=270}
\caption{The relative difference in $B$-band magnitude (normalized at
maximum light) between an unextinguished light curve and several
increasingly extinguished light curves.\label{red_lcs}}
\end{figure}

\clearpage

\begin{figure}[p]
\psfig{file=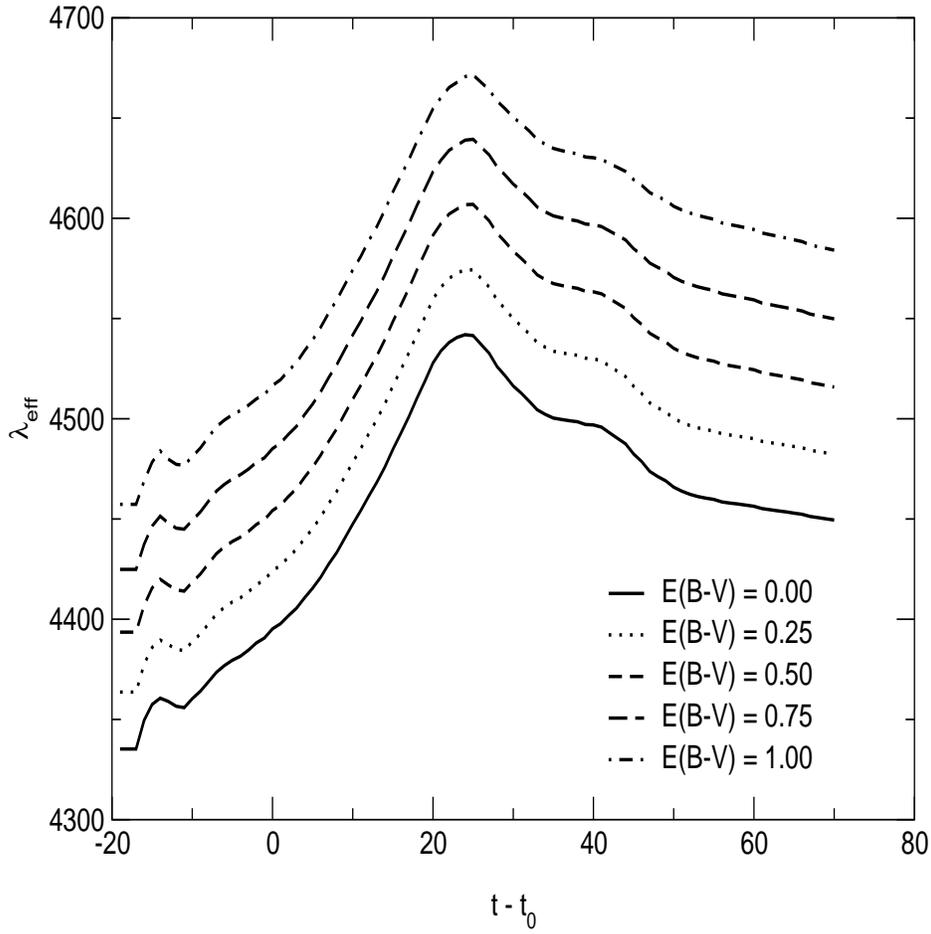,height=6.0in,width=6.0in,angle=270}
\caption{$\lambda_{eff}$ as a function of time for various levels of
extinction in the $B$-band. Note that for a given amount of extinction
a supernova naturally becomes cooler and redder as a function of time
in addition to the expected shift in the effective wavelength as a
function of extinction.\label{lameff}}
\end{figure}

\begin{figure}[p]
\psfig{file=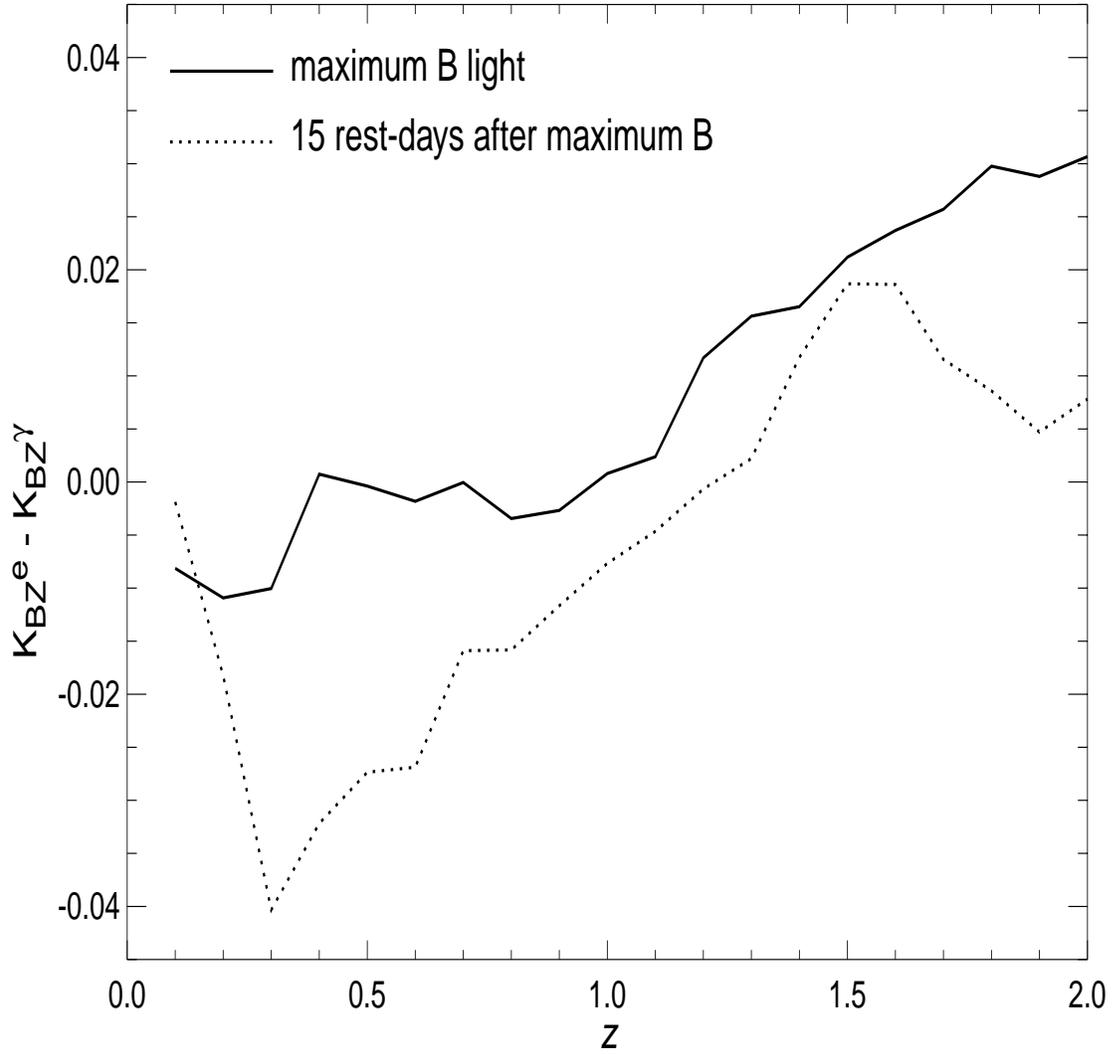,height=6.0in,width=6.0in,angle=0}
\caption[fg1.eps]{$K_{BZ}^\epsilon - K_{BZ}^\gamma$ for a standard Type Ia
supernova at $B$ maximum and 15 days after maximum as a function of
redshift.  Measurements in $I$ and bluer filters for $z<1$ supernovae
and $J$ and redder filters for $z>1.5$ would provide a better match of
observed spectral regions.\label{counts}}
\end{figure}

\end{document}